\documentclass[aps,pre,twocolumn,superscriptaddress]{revtex4-2}
\usepackage{amsmath}
\usepackage{amsfonts}
\usepackage{amssymb}
\usepackage{lmodern,dsfont}
\usepackage{graphicx}
\usepackage[usenames,dvipsnames]{xcolor}
\usepackage{bm}
\usepackage{xr}
\usepackage{siunitx}
\usepackage[english=american]{csquotes}

\newcommand{\kb}{k_\text{B}}
\newcommand{\av}[1]{\langle #1 \rangle}
\newcommand{\Av}[1]{\left\langle #1 \right\rangle}
\newcommand{\nn}{\nonumber \\}
\newcommand{\n}{\nonumber}
\newcommand{\bmc}[1]{\bm{\mathcal{#1}}}

\newcommand{\grad}{\bm{\nabla}}

\renewcommand{\eqref}[1]{Eq.~(\ref{#1})}

\begin{document}

\author{Andreas Dechant}
\affiliation{Department of Physics \#1, Graduate School of Science, Kyoto University, Kyoto 606-8502, Japan}
\author{Jakob H{\"u}pfl}
\affiliation{Institute for Theoretical Physics, Vienna University of Technology (TU Wien), A-1040 Vienna, Austria}
\author{Shuta Kobayashi}
\affiliation{Department of Physics \#1, Graduate School of Science, Kyoto University, Kyoto 606-8502, Japan}
\author{Sosuke Ito}
\affiliation{Universal Biology Institute, The University of Tokyo, 7-3-1 Hongo, Bunkyo-ku, Tokyo 113-0033, Japan}
\author{Stefan Rotter}
\affiliation{Institute for Theoretical Physics, Vienna University of Technology (TU Wien), A-1040 Vienna, Austria}

\title{Precision and cost of feedback cooling}
\date{\today}

\begin{abstract}
We investigate the consequences of information exchange between a system and a measurement-feedback apparatus that cools the system below the environmental temperature.
A quantitative relationship between entropy pumping and information acquired about the system is derived, showing that, independent of the concrete realization of the feedback, the latter exceeds the former by a positive amount of excess information flow.
This excess information flow satisfies a trade-off relation with the precision of the feedback force, which places strong constraints on both the information-theoretic cost of feedback cooling and the required magnitude of the feedback force.
From these constraints, a fundamental lower bound on the energetic cost of optical feedback cooling is derived.
Finally, the results are demonstrated for feedback cooling by coherent light scattering. 
We show that measurement precision is the major factor determining the attainable temperature.
Precise measurements can also be leveraged to reduce the required feedback force, leading to significantly more energy-efficient cooling close to the fundamental bound for realistic parameter values.
\end{abstract}

\maketitle

Feedback has become a common tool for cooling physical systems to fractions of the environmental temperature \cite{Urs03,Bus06,Li11,Gie12} and even to the quantum-mechanical ground state \cite{Guo19,Del20,Kam21}.
By measuring the velocities of a system of particles, it is possible counteract their thermal motion with an appropriate, effectively velocity-dependent, force.
This velocity dependence leads to an effect called entropy pumping \cite{Kim04,Kim07}, which compensates for heat absorption from the environment and allows particles to maintain a kinetic temperature below that of its environment.

From a thermodynamic point of view, feedback cooling crucially relies on the physicality of information:
First, we the velocity of a particle is determined from an observation of its motion, and then, based on the result, an appropriate force is applied to the particle.
This process of measurement and feedback is performed by another physical system, which we refer to as measurement-feedback apparatus, or \textit{feedback controller} for short.
In effect, the feedback controller plays the role of Maxwell's demon, using the information obtained about the system to have the particles perform work against the feedback force, thereby cooling them \cite{Ito11}.
Information thermodynamics \cite{Sag12,Ito13,Har14,Hor14,Par15} quantifies the information flow between the system and the feedback controller, resulting in a modified second law of thermodynamics \cite{Ito11,Mun13,Hor14b,Ros16}.

While both entropy pumping and information thermodynamics have been used to characterize feedback cooling, the connections between these concepts remain largely unexplored, leaving the following questions unanswered:
First, both entropy pumping \cite{Kim04} and information flow \cite{Ito11} are required for feedback cooling, but how are they quantitatively related?
Second, how are entropy pumping and information flow related to the physical feedback force that is acting on the system?
Third, what are the consequences of these relations for concrete implementations of the feedback process?

\begin{figure*}
\includegraphics[width=1\textwidth]{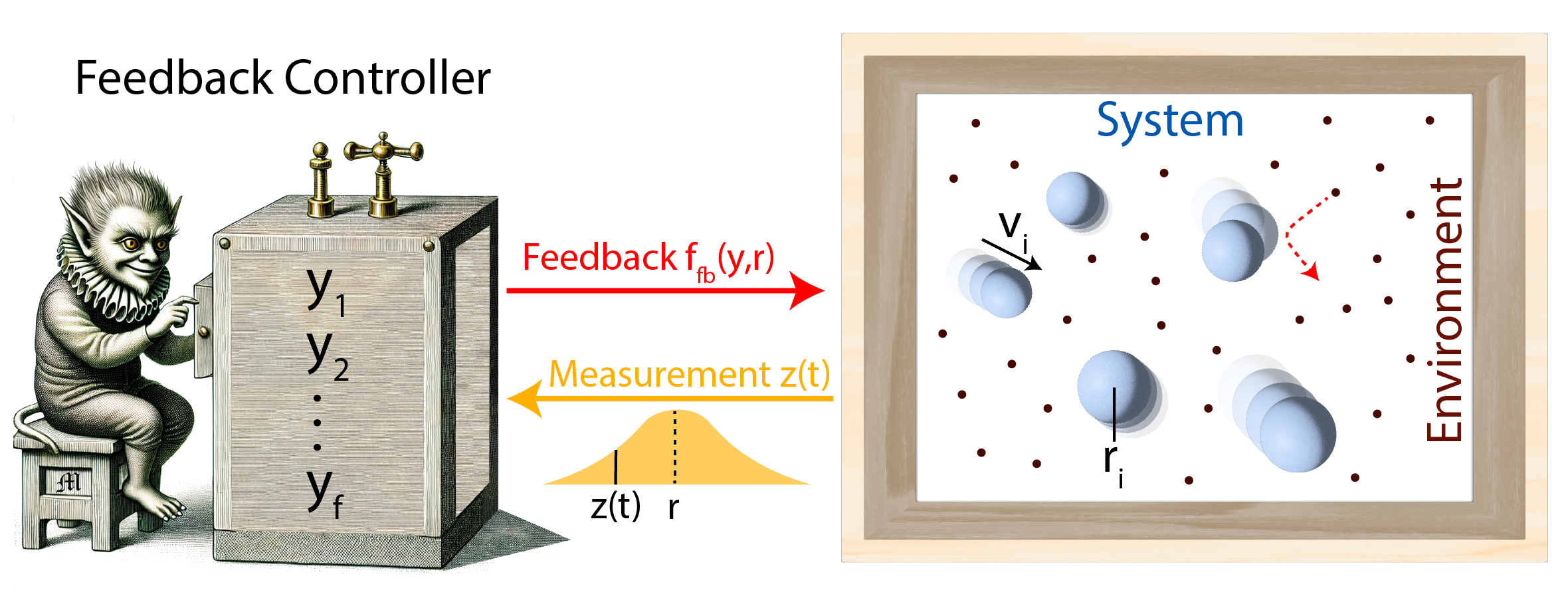}
\caption{Illustration of the feedback cooling setup. A system of particles (blue spheres) with positions $r_i$ and velocities $v_i$ is in thermal equilibrium with the environment. The feedback controller shown on the left collects information on the system through measurements $z(t)$ to update the internal state of the controller (given by $\vec{y}$) and applies a feedback force $\vec{f}_\text{fb}(\vec{y},\vec{r})$.}
\label{fig-illustration}
\end{figure*}

We address these questions by analyzing the information thermodynamics of a system S (as depicted in Fig.~\ref{fig-illustration}) consisting of interacting, underdamped particles (blue spheres) with positions $\bm{r}$ and velocities $\bm{v}$ in contact with a thermal environment.
Based on a measurement of S, a feedback controller F changes its own configuration $\bm{y}$, which determines the feedback force $\bm{f}^\text{fb}(\bm{r},\bm{y})$ applied to the particles.
Since the configuration $\bm{y}$ of F depends on the velocities of the particles through the measurement, the feedback can be also described by an effective velocity-dependent feedback force $\bar{\bm{f}}^\text{fb}(\bm{r},\bm{v})$.
We aim to understand how the velocity dependence of the feedback force is related to the information acquired about the system by the feedback controller and to connect practical implementations of feedback cooling to information thermodynamics.

\section{Main results}

\subsection*{Information flow and entropy pumping}
The first main result is a quantitative relation between the rates of information flow $l^\text{F}$ and entropy pumping $\sigma^\text{epu}$ in the steady state,
\begin{align}
l^\text{F} = \sigma^\text{epu} + l^\text{ex} \label{excess-epu-relation} ,
\end{align}
involving the positive excess information flow $l^\text{ex}$ (see Methods section).
If the temperature of the system is lower than the environmental one, heat continuously flows from the environment to the system at a rate $\dot{Q}^\text{S} < 0$, increasing the system's entropy.
To maintain a steady state at reduced temperature, this must be balanced by an entropy pumping rate $\sigma^\text{epu} > 0$ equal to the rate at which the system's entropy would increase in the absence of the feedback \cite{Kim04,Kim07}.
However, the negative rate of dissipation $\dot{Q}^\text{S} < 0$ also implies a continuously decreasing environmental entropy, apparently violating the second law of thermodynamics.
This contradiction is resolved by considering the information flow from the system to the feedback controller \cite{All09,Har14,Hor14,Hor14b}, measured by the learning rate $l^\text{F}$.
The information flow contributes to the heat dissipation rate $\dot{Q}^\text{F}$ of the feedback controller, $\dot{Q}^\text{F} \geq T l^\text{F}$, resulting in a net increase in environmental entropy as required by the second law. 
\eqref{excess-epu-relation} implies that the information acquired by the feedback controller about the system consists of the entropy pumping rate required for cooling the particles, and the excess information flow $l^\text{ex}$, which does not directly contribute to cooling (thus \enquote{excess}) but instead accounts for the information about the system's velocities encoded in the state of the feedback controller.
While the relation $l^\text{F} \geq \sigma^\text{epu}$ was noted for a simple linear model~\cite{Hor14b},
\eqref{excess-epu-relation} shows that the difference is independent of the model and given by the positive excess information flow
\begin{align}
l^\text{ex}  = \frac{\gamma T}{m^2} \text{tr}\big(\bmc{F}_{v}^{\text{F} \vert \text{S}} \big)  \label{excess-definition} . 
\end{align}
Here, $\gamma$ is the damping constant of the particles, $T$ is the temperature of the environment (setting $\kb = 1$) and $m$ is the mass of a particle.
$\bmc{F}^{\text{F} \vert \text{S}}_v$ is the Fisher information matrix of the conditional probability density $p^{\text{F} \vert \text{S}}(\bm{y} \vert \bm{r},\bm{v})$ of the feedback controller with respect to velocity, $(\bmc{F}^{\text{F} \vert \text{S}}_v)_{jk} = \av{\partial_{v_j} \ln p^{\text{F} \vert \text{S}} \partial_{v_k} \ln p^{\text{F} \vert \text{S}}}$.

The entropy pumping rate can likewise be expressed using the Fisher information matrix $\bmc{F}_{v}^\text{S}$ of the probability density of the system with respect to velocity
\begin{align}
\sigma^\text{epu} = \frac{\gamma T}{m^2} \text{tr} \Big( \bmc{F}_{v}^\text{S} - \frac{m}{T} \bm{I} \Big) \label{epu-fisher},
\end{align}
where $\bm{I}$ denotes the identity matrix.
Note that the second term in the trace, $m \bm{I}/T$, is equal to the Fisher information of the thermal equilibrium distribution.
The entropy pumping rate is positive if the state of the system contains more information about the velocity than a thermal equilibrium state.
The expression \eqref{epu-fisher} implies a bound on the kinetic temperature $T_\text{K} = m \av{\Vert \bm{v} \Vert^2}/d$ of the system (methods),
\begin{align}
\sigma^\text{epu} \geq \frac{\gamma d}{m} \bigg( \frac{T}{T_\text{K}} - 1 \bigg), \label{epu-temperature-bound}
\end{align}
where $d$ is the number of degrees of freedom of the system. 
This explicitly demonstrates that reducing the kinetic temperature below the environmental temperature requires a positive entropy pumping rate.

The Fisher information of the system is related to the conditional Fisher information of the feedback controller by the chain rule,
\begin{align}
\bmc{F}_{v}^{\text{F} + \text{S}} = \bmc{F}_{v}^{\text{F} \vert \text{S}} + \bmc{F}_{v}^{\text{S}} \label{fisher-chain-rule},
\end{align}
where $\bmc{F}_{v}^{\text{F} + \text{S}}$ is the Fisher information of the joint state of system and feedback controller.
Combining this with \eqref{excess-epu-relation} and \eqref{epu-fisher}, we can express the information flow as
\begin{align}
l^\text{F} = \frac{\gamma T}{m^2} \text{tr} \Big( \bmc{F}_{v}^\text{S+F} - \frac{m}{T} \bm{I} \Big).
\end{align}
Therefore, the learning rate accounts for the information about the velocity contained in the state of both system and feedback controller, while the excess information flow \eqref{excess-definition} measures the additional information contained in the state of the feedback controller.


\subsection*{Precision-dissipation trade-off}
Since the excess information flow increases the dissipation of the feedback controller, one might seek to eliminate it.
This is prohibited by our second main result, which is a trade-off relation between entropy pumping, precision and excess information flow,
\begin{align}
\gamma T \big(\sigma^\text{epu} \big)^2 \leq l^\text{ex} \Av{ \Vert \bm{f}^\text{fb} - \bar{\bm{f}}^\text{fb} \Vert^2} = l^\text{ex} \Av{\Vert \delta \bm{f}^\text{fb} \Vert^2} \label{excess-epu-tradeoff} .
\end{align}
The effective, velocity-dependent feedback force $\bar{\bm{f}}^\text{fb}(\bm{r},\bm{v}) = \int d\bm{y} \bm{f}^\text{fb}(\bm{r},\bm{y}) p^{\text{F} \vert \text{S}}(\bm{y} \vert \bm{r},\bm{v})$ is the conditional average of the feedback force. 
The effective force $\bar{\bm{f}}^\text{fb}$ describes the average effect of the feedback on the system and determines the entropy pumping rate (see Methods),
\begin{align}
\sigma^\text{epu} = -\frac{1}{m} \Av{\grad_v \cdot \bar{\bm{f}}^\text{fb}} . \label{epu-definition}
\end{align}
However, the actual feedback force applied to the system differs from the effective one by $\delta \bm{f}^\text{fb} = \bm{f}^\text{fb} - \bar{\bm{f}}^\text{fb}$ due to the finite precision of the measurement and feedback process.
The trade-off relation \eqref{excess-epu-tradeoff} between information flow and the feedback precision imposes a strong constraint on any physical implementation of feedback cooling:
In order to reproduce a desired velocity-dependent force with perfect precision ($\delta \bm{f}^\text{fb} = 0$), the rate of information transfer from the system to the feedback controller would have to diverge.

\subsection*{Feedback efficiency}
The efficiency of the feedback process can be characterized in different ways.
In Ref.~\cite{Hor14} the information-thermodynamic efficiencies were defined as
\begin{gather}
\eta^\text{td,S} = \frac{- \dot{Q}^\text{S}}{T l^\text{F}} \quad \text{and} \quad \eta^\text{td,F} = \frac{T l^\text{F}}{\dot{Q}^\text{F}} ,
\end{gather}
which measure how effectively the information flow is converted into a negative dissipation of the system, and how efficiently the dissipation of the feedback controller is converted into information flow.
On the other hand, \eqref{excess-epu-relation} allows defining the entropy pumping efficiency
\begin{align}
\eta^\text{epu} = \frac{\sigma^\text{epu}}{l^\text{F}} = \frac{\sigma^\text{epu}}{\sigma^\text{epu} + l^\text{ex}} \label{epu-efficiency} ,
\end{align}
which measures how effectively the information flow is converted into entropy pumping.
Moreover, \eqref{epu-temperature-bound} suggests the definition of a cooling efficiency
\begin{align}
\eta^\text{cool} = \frac{\gamma d \big( \frac{T}{T_\text{K}} - 1 \big)}{m \sigma^\text{epu}}, \label{cooling-efficiency}
\end{align}
which measures how well entropy pumping is converted into a reduction in the kinetic temperature.
In the cooling regime, all efficiencies satisfy $0 \leq \eta \leq 1$.
The cooling efficiency and entropy pumping efficiency are related to the information-thermodynamic efficiency by
\begin{align}
\eta^\text{td,S} = \frac{1 - \frac{T_\text{K}}{T}}{\frac{T}{T_\text{K}} - 1} \eta^\text{cool} \eta^\text{epu} .
\end{align}
Thus, perfect conversion of information into heat flow can only be achieved for negligible cooling $T \approx T_\text{K}$.
However, rather than heat flow, the utility of feedback cooling is the reduction of the kinetic temperature, which suggests using $\eta^\text{epu}$ and $\eta^\text{cool}$ as the relevant measures of efficiency.
The feedback efficiency can then be defined as
\begin{align}
\eta^\text{fb} = \eta^\text{cool} \eta^\text{epu} \eta^\text{td,F} = \frac{T \frac{\gamma d}{m} \big( \frac{T}{T_\text{K}} - 1 \big)}{\dot{Q}^\text{F}},
\end{align}
which relates the reduction of the kinetic temperature to the thermodynamic cost of the feedback controller.
The cooling efficiency attains its maximal value $\eta^\text{cool} = 1$ for an effective feedback force that is linear in the velocity $\bar{\bm{f}}^\text{fb}(\bm{v}) = - \gamma^\text{fb} \bm{v}$ (methods).
Due to \eqref{excess-epu-tradeoff}, the entropy pumping efficiency is bounded,
\begin{align}
\eta^\text{epu} \leq \frac{1}{1 + \frac{\gamma T \sigma^\text{epu}}{\Av{\Vert \delta \bm{f}^\text{fb} \Vert^2}}},
\end{align}
and perfect conversion of information flow into entropy pumping can only be achieved in the limit $\av{\Vert \delta \bm{f}^\text{fb} \Vert^2} \rightarrow \infty$, which requires applying an infinitely strong feedback force to the system.
Finally, the energetic cost and the efficiency $\eta^\text{td,F}$ depend on the concrete implementation of the feedback controller. 
Nevertheless, as discussed in the following, general bounds on these quantities can be derived.


\begin{figure}
\includegraphics[width=.47\textwidth]{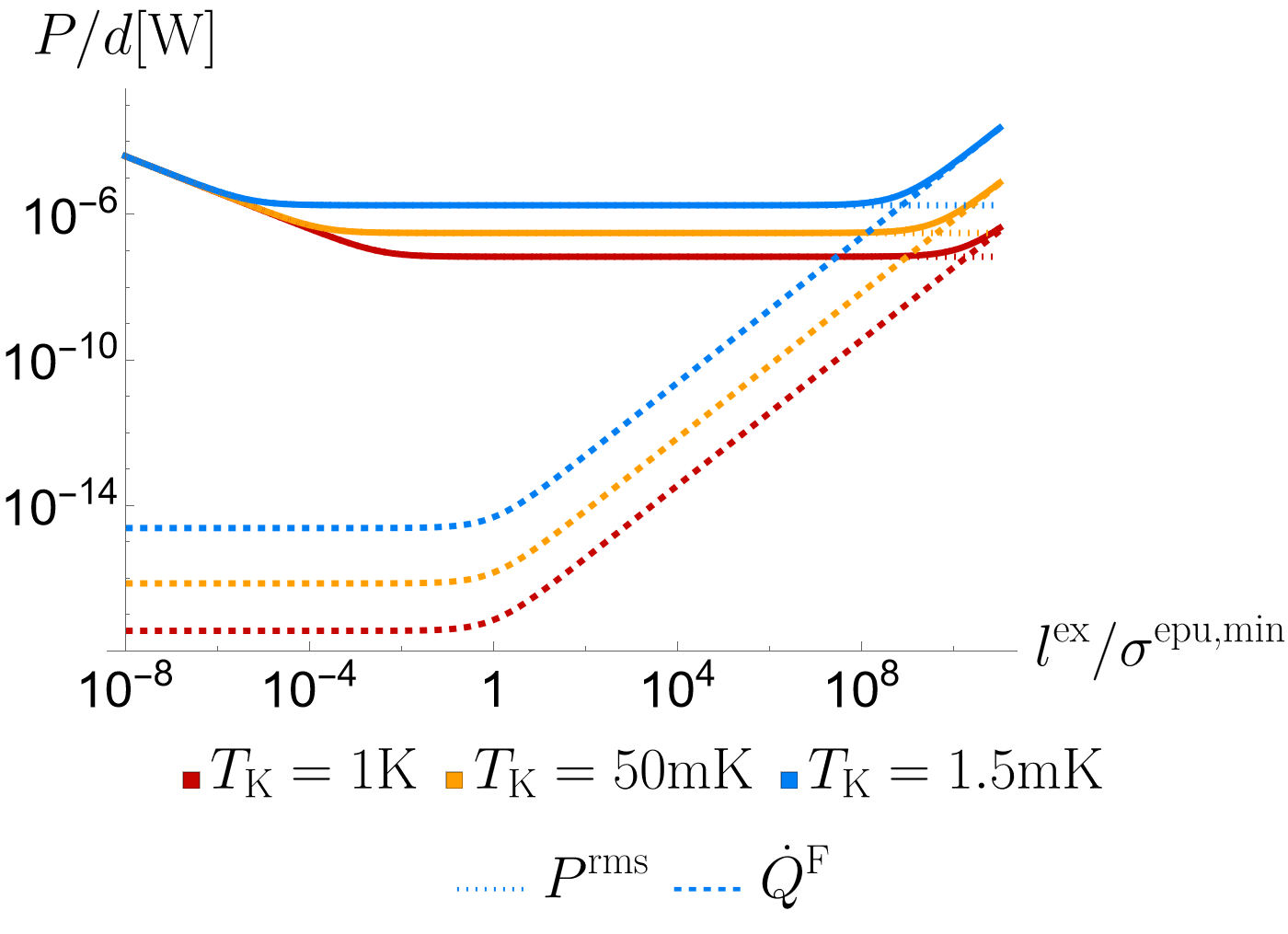}
\caption{The lower bound on the power consumption of optical feedback cooling as a function of the excess information flow for three different kinetic temperatures. 
$l^\text{ex}$ is expressed in units of the minimal entropy pumping rate at a given kinetic temperature, $\sigma^\text{epu,min} = \gamma d/m (T/T_\text{K} -1)$, see \eqref{epu-temperature-bound}. 
Parameters are $T = 300 \text{K}$, $m = \num{3.4e-14} \text{kg}$, $\gamma/(2 \pi m) = 0.46 \si{Hz}$. 
The thermodynamic cost (dashed, \eqref{dissipation-bound}) of the feedback cooling only becomes relevant compared to the power required by the laser (dotted, obtained from \eqref{force-bound}) for very large $l^\text{ex}$.
For small $l^\text{ex}$, on the other hand, the deviations of the cooling force from its effective value become dominant, and we need increasingly larger laser power to maintain the same effective force.
The plateau in between these limits provides a universal lower bound on the cost of optical feedback cooling.}
\label{fig-power-bound}
\end{figure}

\subsection*{Cost of feedback cooling}
Our third main result is that the excess information $l^\text{ex}$ flow provides lower bounds on both the rate of heat dissipation $\dot{Q}^\text{F}$ of the feedback controller and on the magnitude of the feedback force $\av{\Vert \bm{f}^\text{fb} \Vert^2}$ (methods),
\begin{subequations}
\begin{align}
\dot{Q}^\text{F} &\geq \frac{(f^\text{env})^2}{\gamma} \bigg( \frac{T}{T_\text{K}} - 1 + \frac{m l^\text{ex}}{\gamma d} \bigg) \label{dissipation-bound} \\
\Av{\Vert \bm{f}^\text{fb} \Vert^2} &\geq \big( f^\text{env} \big)^2 \bigg( \frac{T}{T_\text{K}} + \frac{T_\text{K}}{T} - 2 \bigg) \bigg( 1 + \frac{T}{T_\text{K}} \frac{\gamma d}{m l^\text{ex}} \bigg) \label{force-bound} .
\end{align} \label{dissipation-force-bounds}%
\end{subequations}
Here, $f^\text{env} = \gamma \sqrt{T d/m}$ is the magnitude of the environmental friction at thermal velocity.
Both the dissipation and the feedback force increase as the kinetic temperature decreases.
Bounds on the required dissipation for a given non-equilibrium process have been a central subject in the recent literature \cite{Bar15,Gin16,Pie18,Dec20,Hor20}. 
Their importance stems from the fact that the heat dissipated by the feedback controller must be compensated by external energy input and thus \eqref{dissipation-bound} quantifies the energetic cost of the feedback process.

By contrast, \eqref{force-bound} is conceptually different from these results, providing a bound on the required force rather than the dissipation.
To illustrate its significance, we focus on optical feedback cooling, where the feedback force is due to interaction between the particles and the surrounding light field, usually from a laser.
Under ideal circumstances (all light is reflected and the incidence vector of the light field is parallel to the force) and in free space, the force is related to the laser power by
\begin{align}
P = \Vert \bm{f}^\text{fb} \Vert \frac{c}{2} ,
\end{align}
where $c$ is the speed of light.
Using this, \eqref{force-bound} provides a bound on the root-mean-square power of the laser, $P^\text{rms} = \sqrt{\av{P^2}}$.
This operational cost occurs in addition to the thermodynamic dissipation, and therefore the overall energetic cost of the feedback process is bounded by $P^\text{tot} \geq \dot{Q}^\text{F} + P^\text{rms}$.
Consequently, both the operational and thermodynamic cost should be factored into the feedback efficiency, which becomes
\begin{align}
\eta^\text{fb} = \frac{T \frac{\gamma d}{m} \big( \frac{T}{T_\text{K}} - 1 \big)}{\dot{Q}^\text{F} + P^\text{rms}} \label{feedback-efficiency}.
\end{align}
Since the bound on $P^\text{rms}$ decreases with the excess information flow, while the bound on $\dot{Q}^\text{F}$ increases, there exists an optimal value of $l^\text{ex}$ that minimizes the bound on the total power.
This minimal value is independent of the concrete value of $l^\text{ex}$ and therefore provides a fundamental bound on the power requirement of optical feedback cooling and thus the feedback efficiency.

As a concrete example, consider $N$ particles in 3 dimensions ($d = 3N$) at room temperature $T = 300 \si{K}$.
Each particle has a mass of $\num{3.2e-14} \si{kg}$ and the thermal damping rate is $\gamma/(2 \pi m) = 0.46 \si{Hz}$, which are typical parameters for optical feedback cooling \cite{Li11}.
Fig.~\ref{fig-power-bound} shows the resulting bound on the power as a function of the excess information flow $l^\text{ex}$. 
The overall lower bound on the power evaluates to $P^\text{tot} \geq 0.0018 \si{mW}$ for cooling to a kinetic temperature of $1.5 \si{mK}$.
This result is about three orders of magnitude smaller than the typical power of the cooling laser ($1 \si{mW}$) used in experiments \cite{Li11}, which, however, differ from the ideal conditions assumed above.
Specifically, since the particles are typically smaller than the beam size, only a fraction of the laser light actually interacts with them, and the force on the particles often arises from the power difference between two or more laser beams, rather than from a single beam.
On the other hand, the bound shown in Fig.~\ref{fig-power-bound} is seven orders of magnitude larger than the power associated with entropy pumping, $T \sigma^\text{epu} \approx 10^{-15} \si{W}$, which clearly demonstrates that for optical cooling, the cost of generating the feedback force dominates the thermodynamic cost.

\subsection*{Direct and indirect degrees of freedom}
In the above discussion, the entire process of measuring the system, computing the corresponding feedback force and applying it to the system was modeled by the degrees of freedom $\bm{y}$ of an unspecified measurement-feedback apparatus.
For example, optical feedback cooling requires detectors that measure the light scattered by particles as well as circuitry that translates the result of the measurement into information about the position of the particle and then uses this information to adjust the power and phase of the cooling laser accordingly.
Ultimately, however, the feedback force is determined by the parameters of the cooling laser (denoted as \enquote{direct} degrees of freedom $\bm{y}^\text{d}$), whereas the \enquote{indirect} degrees of freedom $\bm{y}^\text{i}$ only influence the force via their impact on $\bm{y}^\text{d}$.
The feedback force can thus be written as $\bm{f}^\text{fb}(\bm{r},\bm{y}) = \bm{f}^\text{fb}(\bm{r},\bm{y}^\text{d})$.
This corresponds to separating the feedback controller into two separate physical systems: the first component, labeled $\text{Fd}$, involves the direct degrees of freedom and the second component, labeled $\text{Fi}$, is composed of the indirect ones.
The chain rule for the Fisher information \eqref{fisher-chain-rule} lets us separate the Fisher information as follows
\begin{align}
\bmc{F}_{v}^{\text{Fi} + \text{Fd} + \text{S}} = \bmc{F}_{v}^{\text{Fi} \vert \text{Fd} + \text{S}} + \bmc{F}_{v}^{\text{Fd} \vert \text{S}} + \bmc{F}_{v}^{\text{S}} .
\end{align}
The first term on the right-hand side measures the information about the particles' velocities contained in Fi that is contained neither in Fd nor in S; it provides a positive contribution to the excess learning rate \eqref{excess-definition} and thus to the information flow between the system and the feedback controller,
\begin{align}
l^\text{ex} = l^\text{ex,Fi} + l^\text{ex,Fd} \quad \text{with} \quad l^\text{ex,Fd} = \frac{\gamma T}{m^2} \text{tr}\big(\bmc{F}_{v}^{\text{Fd} \vert \text{S}} \big).
\end{align}
However, since the feedback force does not directly depend on $\bm{y}^\text{i}$, \eqref{excess-epu-tradeoff} remains valid when replacing $l^\text{ex}$ by $l^\text{ex,Fd}$, which therefore provides a tighter bound.
Similarly, $l^\text{ex,Fd}$ provides a tighter bound on the magnitude of the feedback force in \eqref{force-bound}.
Conversely, including the information flow due to the indirect degrees of freedom $l^\text{ex,Fi}$ yields a tighter bound on the dissipated heat \eqref{dissipation-bound}.
This implies that the bound on the feedback force \eqref{force-bound} is not only practically relevant, it is also easier to evaluate in practice, since only the degrees of freedom of the feedback controller that directly contribute to the force have to be modeled.
By contrast, the estimate of the dissipation \eqref{dissipation-bound} becomes more accurate the more detailed the model of the feedback controller is.

\section{Cooling by coherent scattering in one dimension.}
We now discuss the consequences of the above results in a simple and concrete, yet realistic model of optical feedback cooling.
The model is based on \cite{Mun13}, but explicitly models the noisy measurement process and accounts for the generation of the feedback force via optical scattering and gradient forces.

\subsection*{Scattering forces}
In practice, many feedback cooling schemes, such as for levitated particles, rely on optomechanical effects induced by the interaction between a light field and the particles.
Focusing on the simple case of an underdamped single particle trapped in a one-dimensional harmonic potential, the interaction force can be written as \cite{Amb17,Hüpfl2023b,Hup23}
\begin{align}
f^\text{fb}(r,c_{l,\text{in}},c_{r,\text{in}}) = \frac{1}{2 \omega} \begin{pmatrix} c_{l,\text{in}}^* \\ c_{r,\text{in}}^* \end{pmatrix}^\text{T} \bm{Q}_r(r) \begin{pmatrix} c_{l,\text{in}} \\ c_{r,\text{in}} \end{pmatrix} \label{optical-force}.
\end{align}
Here, $c_{l,\text{in}}$/$c_{r,\text{in}}$ are the complex amplitudes of an incoming plane-wave electrical field from the left/right and $\omega$ is its frequency.
$\bm{Q}_r(r)$ is the generalized Wigner-Smith operator with respect to the particle's position $r$
\begin{align}
\bm{Q}_r(r) = - i \bm{S}^\dagger(r) \partial_r \bm{S}(r) ,
\end{align}
where $\bm{S}(r)$ is the \enquote{frozen} scattering matrix that relates the incoming field to the outgoing field by $(c_{l,\text{out}},c_{r,\text{out}}) = \bm{S}(r) (c_{l,\text{in}},c_{r,\text{in}})$.
The feedback controller now has the task of choosing the incoming amplitudes $c_{l,\text{in}}$ and $c_{r,\text{in}}$ based on the measurement of $r$ and $v$ in a way that reduces the steady-state kinetic temperature $T_\text{K} = m \av{v^2}$.

\subsection*{Measurement and feedback}
The measurement and feedback process can be approximated by a solvable, linear model similar to Ref.~\cite{Mun13}, by introducing the dynamical variable
\begin{align}
z(t) &= \frac{1}{\tau} \int_0^t dt \ e^{-\frac{t-s}{\tau}} \big( r(s) + \sqrt{2 D} \eta(s) \big) \label{measurement-equations} .
\end{align}
$z$ corresponds to an exponentially weighted average of the particle position $r$ with additional measurement noise $\eta$ with magnitude $D$.
We assume that $\eta(t)$ is Gaussian noise with correlation time $\tau^\text{noise}$, that is $\av{\eta(t) \eta(t')} = \exp[-|t-t'|/\tau^\text{noise}]/(2 \tau^\text{noise})$.
In the limit $\tau^\text{noise} \rightarrow 0$ this recovers Gaussian white noise; the joint system S+F becomes non-bipartite in this limit and the finite correlation time is included to avoid the ensuing issues \cite{Che19} (see Appendix C) for details).
\eqref{measurement-equations} accounts for the fact that the instantaneous position of the particle cannot be determined; instead, the measurement signal (for example the light scattered by the particle) must be integrated over a certain interval $\tau$.
From the dynamics of $z$, the velocity of the particle is estimated as $u = \dot{z}$.
The internal degrees of freedom of the feedback controller are then $\bm{y} = (z,u)$, in terms of which the amplitudes of the incoming light field are parameterized as
\begin{align}
c_{l,\text{in}} = \sqrt{\frac{P_0}{v_0} \vert u \vert} \Theta(-u), \qquad c_{r,\text{in}} = \sqrt{\frac{P_0}{v_0} \vert u \vert} \Theta(u) \label{light-modes}
\end{align}
where $P_0$ is a constant with dimensions of power, $v_0$ is a constant with dimensions of velocity, whose value does not impact the final result and is therefore set to $v_0 = v_\text{th} = \sqrt{T/m}$, and $\Theta(u)$ denotes the Heaviside function.
The light field resulting from \eqref{light-modes} has an intensity proportional to the measured velocity $u$ and is incoming from the left (right) when the particle is moving to the left (right), giving rise to the following feedback force (see Methods)
\begin{align}
f^\text{fb}(u) &= - \gamma^\text{fb} u \qquad \text{with} \label{feedback-force} \\
\gamma^\text{fb} &= \frac{P_0}{c v_0} \frac{4 (1 - \epsilon_r)^2 \sin(q L)^2}{4 \epsilon_r + (1 - \epsilon_r)^2 \sin(q L)^2} \n ,
\end{align}
with the feedback damping $\gamma^\text{fb}$.
Here, $\epsilon_r$ is the relative dielectric permittivity of the particle, $L$ its diameter and $q = \sqrt{\epsilon_r} k$ with $k = c/\omega$ the wave vector of the incoming light field.
The above parameterization of the light field assumes that the dominant error stems from the measurement of the velocity, and that generating the feedback force from the result of the measurement incurs no further errors.
If the trapping potential is parabolic, $U(r) = \kappa r^2/2$, then the joint equations of motion for the particle and feedback controller are linear, 
\begin{align}
\dot{r} = v, \quad m \dot{v} &= -\kappa r - \gamma v - \gamma^\text{fb} u + \sqrt{2 \gamma T} \xi, \nn
\tau \dot{z} &= - (z - r) + \sqrt{2 D} \eta \\
\tau^\text{noise} \dot{\eta} &= - \eta + \tilde{\xi}, \n
\end{align}
where we introduced the auxiliary Gaussian white noise $\tilde{\xi}$.
The linearity of the equations of motions allows solving for the steady state probability density, which is Gaussian, and calculating all relevant quantities analytically.

\begin{figure}
\includegraphics[width=0.47\textwidth]{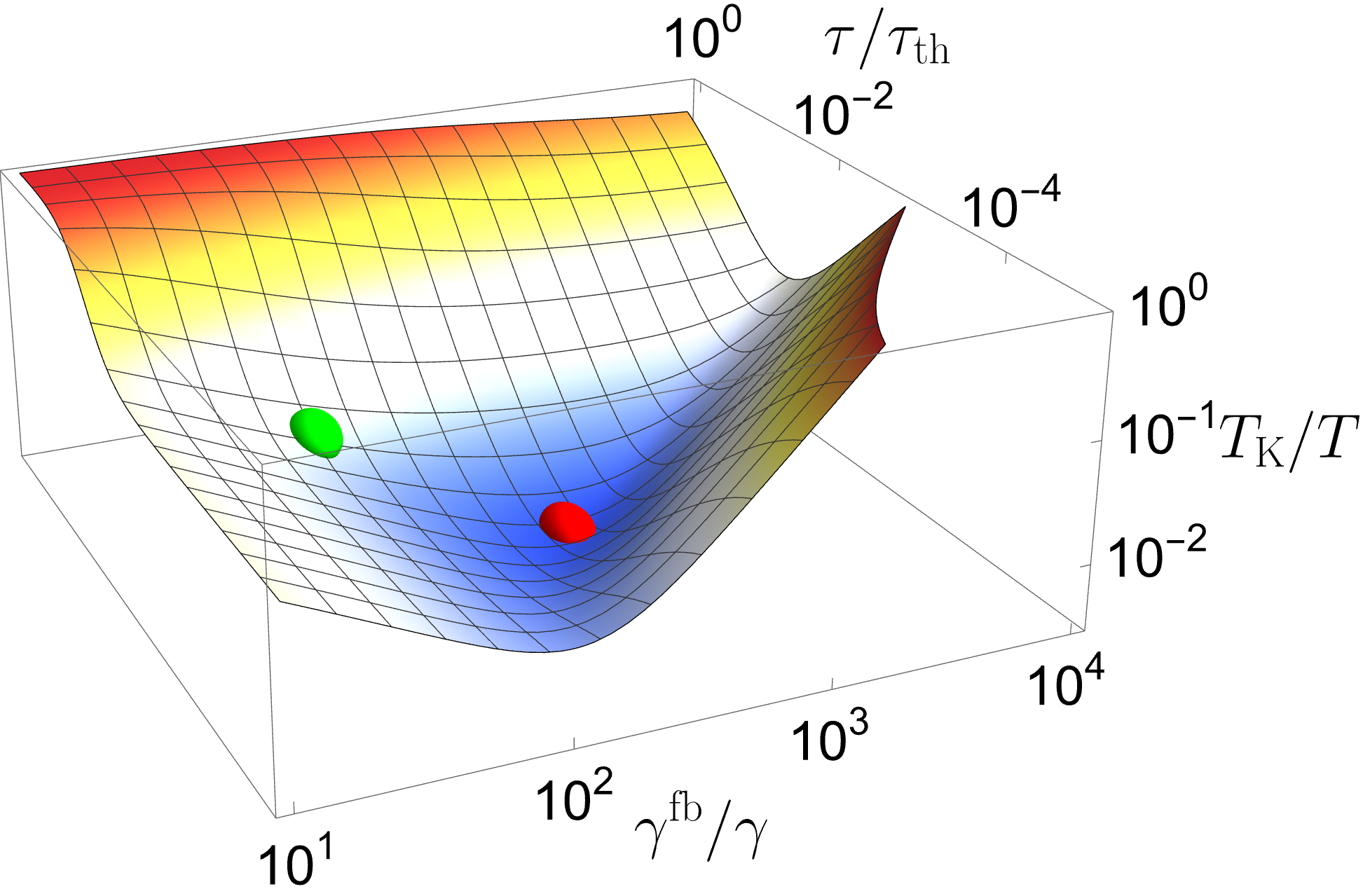}
\caption{Kinetic temperature as a function of the integration time $\tau$ of the feedback controller (in units of the thermal relaxation time $\tau_\text{th} = \gamma/m$) and the feedback damping rate $\gamma^\text{fb}$ (in units of the environmental damping $\gamma$). The measurement noise is $D = \num{1.9e-7} \av{r^2}^\text{th} \tau^\text{osc}$, where $\av{r^2}^\text{th} = T/\kappa$ is the magnitude of equilibrium position variance and $\tau^\text{osc} = 2\pi \sqrt{m/\kappa}$ is the period of the oscillations in the harmonic trapping potential. The kinetic temperature shows a global minimum of $T_\text{K} = 0.005 T$, indicated by a red dot. Maximizing the feedback efficiency \eqref{feedback-efficiency} reduces the required power from $1.2 \si{mW}$ to $0.042 \si{mW}$ but increases the kinetic temperature to $T_\text{K} = 0.026 T$ (green dot).} 
\label{fig-temperature}
\end{figure}

\begin{figure}
\includegraphics[width=0.47\textwidth]{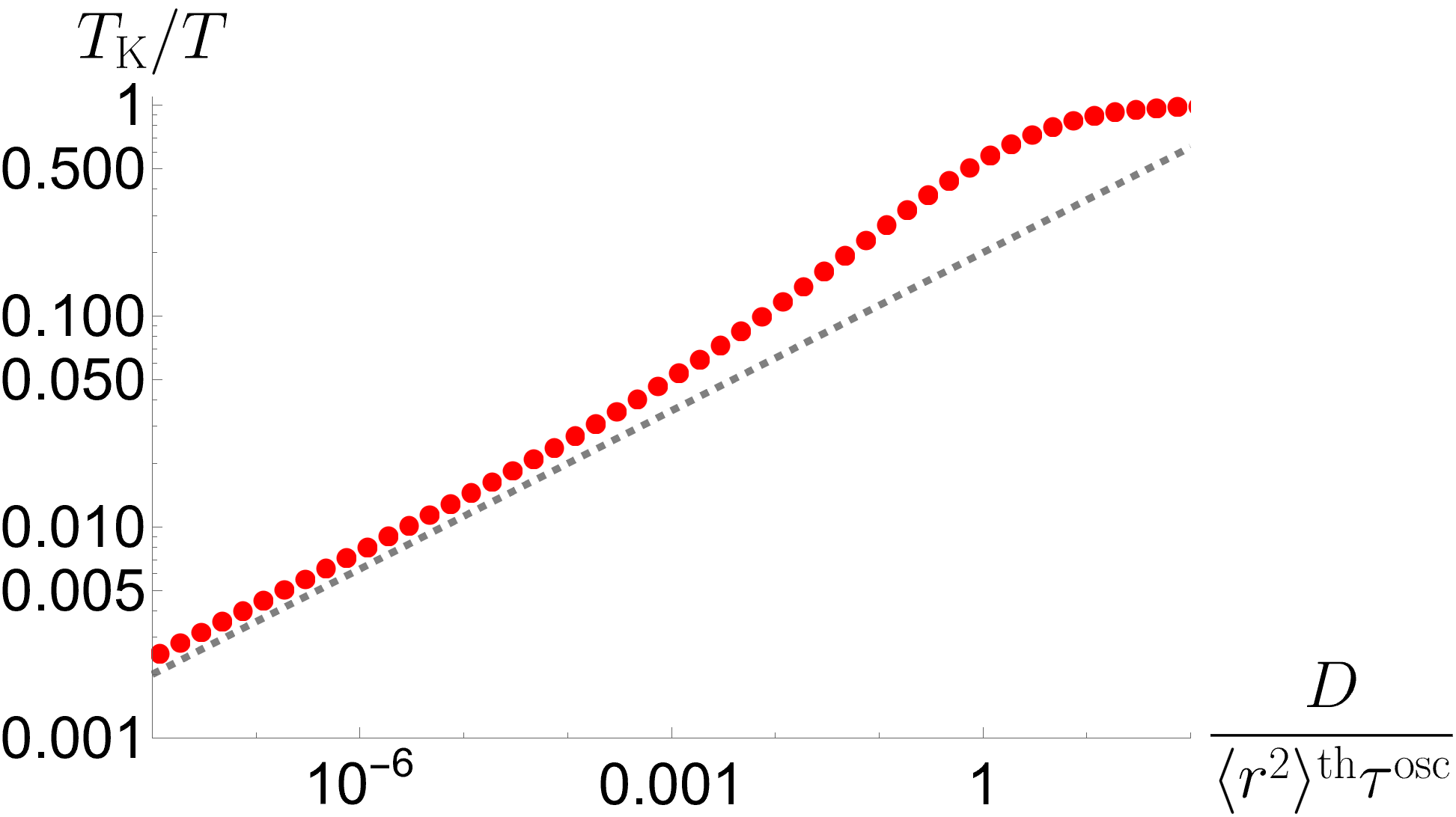}
\caption{The minimal kinetic temperature $T_K$ as a function of the magnitude of the measurement noise. The red circles show the value obtained by numerically minimizing $T_\text{K}$ with respect to the integration time $\tau$ and the magnitude of the feedback force $\gamma^\text{fb}$. For large measurement noise, no cooling is possible, since no useful information about the system's velocity can be obtained from the measurement. The dashed line shows the asymptotic proportionality $T_\text{K} \propto (D)^{1/4}$ for small measurement noise.} 
\label{fig-temperature-noise}
\end{figure}

\subsection*{Kinetic temperature}
Fig.~\ref{fig-temperature} shows the kinetic temperature of the particle depending on the integration time $\tau$ of the feedback controller and the feedback damping $\gamma^\text{fb}$ that is proportional to the overall power $P_0$ of the light fields.
For given values of the remaining parameters (methods), the kinetic temperature attains a global minimum for specific values of $\tau$ and $\gamma^\text{fb}$.
If the integration time $\tau$ is too short, then the signal will be dominated by measurement noise; if it is too long, then $z$ does not capture the instantaneous position of the particle; in both cases, no effective cooling is possible.
It is also clear that there is no cooling in the limit $P_0 \rightarrow 0$ where the feedback force vanishes.
On the other hand, when the light fields are too strong, the measurement noise will be amplified and fed back into the system, effectively heating it.
The values of $\tau$ and $\gamma^\text{fb}$ minimizing the kinetic temperature are determined by numerical optimization of the analytic expression for the kinetic temperature (see Appendix C for details).

The crucial parameter determining the minimal value of the kinetic temperature is the magnitude of the measurement noise $D$, see Fig.~\ref{fig-temperature-noise}.
As the measurement noise tends to zero, so does the attainable kinetic temperature:
For perfect measurement and feedback, the system could in principle be cooled to zero temperature by applying infinitely strong feedback forces.
Numerically, the scaling $T_\text{K}^\text{min} \propto (D^\text{mn})^{1/4}$ is observed; thus, reducing the kinetic temperature by one order of magnitude requires four orders of magnitude of reduction in the measurement noise.
This highlights the critical role of precise measurements for feedback cooling.


\begin{figure}
\includegraphics[width=0.47\textwidth]{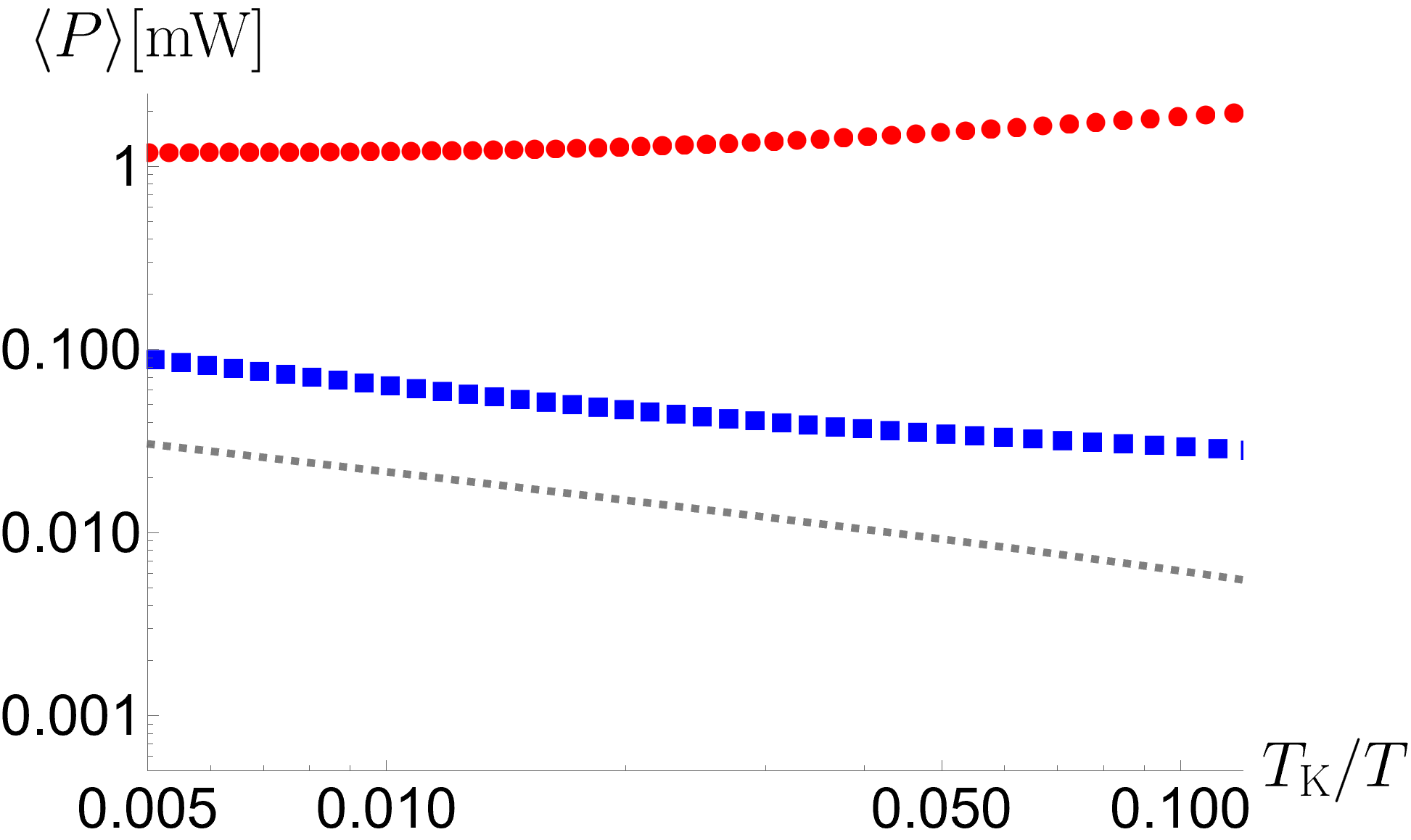}
\caption{The average laser power \eqref{laser-power} as a function of the kinetic temperature. The red circles are obtained by minimizing the kinetic temperature irrespective of the power, the blue squares are obtained by maximizing the efficiency \eqref{feedback-efficiency}. Prioritizing the efficiency reduces the laser power for a given temperature by more than one order of magnitude, coming within a factor $3$ of the universal bound obtained from \eqref{dissipation-force-bounds} (gray dashed line), corresponding to the minimal value of the power in Fig.~\ref{fig-power-bound}.} 
\label{fig-power}
\end{figure}

\subsection*{Maximal feedback efficiency}
To quantify the energetic cost of the feedback, the laser power is obtained from \eqref{light-modes} as
\begin{align}
P^\text{rms} = \sqrt{\Av{(\vert c_{l,\text{in}} \vert^2 + \vert c_{r,\text{in}} \vert^2)^2}} = P_0 \frac{ \sqrt{\av{u^2}}}{v_0} \label{laser-power} .
\end{align}
Minimizing the kinetic temperature corresponds to maximizing the numerator of the feedback efficiency in \eqref{feedback-efficiency}.
By contrast, as shown in Fig.~\ref{fig-power}, the parameters maximizing the feedback efficiency result in a significantly lower power at a given kinetic temperature. 
Since the kinetic temperature is determined only by the effective feedback force $\bar{\bm{f}}^\text{fb}$, minimizing the former corresponds to maximizing the latter without regard to the actual feedback force $\bm{f}^\text{fb}$.
However, as argued above, the energetic cost of the feedback is dominated by the laser power, which is proportional to the actual force applied to the particles.
Reducing this force while maintaining the effective feedback force requires more precise feedback, which, in turn requires a larger information flow.
Despite the fact that this leads to more dissipation from a thermodynamic point of view, this is offset by the reduced laser power, resulting in more efficient feedback overall.

\section{Discussion}
Effective, velocity-dependent forces used in feedback cooling have to be generated by a physical system performing the feedback.
\eqref{excess-epu-relation} relates the information flow between the system and the feedback controller to entropy pumping.
In particular, the mere presence of a physical feedback controller causes the information flow to exceed the amount of entropy pumping, leading to additional thermodynamic cost for the feedback in the form of entropy production.
In principle, one might be tempted to eliminate this additional dissipation, however, the trade-off \eqref{excess-epu-tradeoff} implies that this can only be achieved at the cost of infinitely strong feedback forces.

The result \eqref{excess-epu-relation} also applies to feedback processes whose goal is not to cool particles but rather to extract work from the environmental fluctuations.
Any apparent violation of the second law is bounded by the entropy pumping rate $\dot{S}_t^\text{S} - \dot{Q}_t^\text{S}/T \geq - \sigma_t^\text{epu}$.
As a consequence, the trade-off \eqref{excess-epu-tradeoff} also bounds the extent of the violation of the conventional second law.
Just as in the case of feedback cooling, such a violation entails an excess information flow, which can only be made to vanish in the limit of infinitely strong feedback forces.
It would be interesting to investigate how these quantities behave in the overdamped limit, where no feedback cooling is possible, and the feedback controller's dependence on the system's velocity is expected to vanish to leading order.

Bounds on the thermodynamic cost of non-equilibrium processes have been the topic of many studies over recent years \cite{Bar15,Gin16,Pie18,Dec20,Hor20}, including for systems involving feedback and information flow \cite{Pot19,Wol20,Van20b,Ots20,Nak21,Tan23}.
For optical feedback, the operational cost of generating the feedback force can dominate the thermodynamic cost of information flow; both need to be taken into account to realize efficient feedback.
Moreover, while the physical mechanism behind the feedback force can vary, in many experimental settings, the ability to generate large and precise feedback forces is a serious limiting factor.
Therefore, bounds such as \eqref{excess-epu-tradeoff} and \eqref{dissipation-force-bounds} that reveal the magnitude and precision of the force necessary to achieve a desired effect are thus potentially of great practical relevance.

\section{Methods}
\subsection*{Theoretical description of feedback cooling}
We consider feedback cooling of a (generally interacting) system consisting of $n$ particles, whose positions and velocities we summarize in the ($d = 3 n$)-dimensional vectors $\bm{r}$ and $\bm{v}$.
The particles are in contact with a thermal environment at temperature $T$ and, for simplicity, we assume that all particles have the same mass $m$ and damping coefficient $\gamma$.
Then, the motion of the particles can be described by the underdamped Langevin equations \cite{Cof17}
\begin{align}
\dot{\bm{r}} &= \bm{v} \label{langevin} \\
m \dot{\bm{v}} &= - \grad_r U(\bm{r}) - \gamma \bm{v} + \bm{f}^\text{fb}(\bm{r},\bm{y}) + \sqrt{2 \gamma T} \bm{\xi} \n,
\end{align}
where $U(\bm{r})$ is a potential containing external and interaction forces and $\bm{\xi}$ is a vector of mutually independent Gaussian white noises describing the thermal fluctuations.
The feedback force $\bm{f}^\text{fb}(\bm{r},\bm{y})$ is determined by the current configuration of the feedback controller, which is represented as the vector $\bm{y} \in \mathbb{R}^f$.
The feedback force can depend on the positions of the particles (as is used in parametric feedback cooling), but is assumed not to depend explicitly on the velocities.
The task of the feedback controller is to determine $\bm{y}$ in such a way as to reduce the fluctuations of the system, which are quantified by the kinetic temperature $T_\text{K} = m \av{\Vert \bm{v} \Vert^2}/d$, where $\Av{\ldots}$ denotes an average over realizations of the noise.

Equivalent to the stochastic description \eqref{langevin}, the time-evolution of the phase-space probability density $p_t^{\text{S} + \text{F}}(\bm{r},\bm{v},\bm{y})$ obeys the continuity equation \cite{Ris86}
\begin{align}
\partial_t p^{\text{S} + \text{F}}_t = - \grad_r  \cdot \bm{j}^r_{t} - \grad_v \cdot \bm{j}^v_{t} - \grad_y \cdot \bm{j}^y_{t} \label{continuity} ,
\end{align}
where the arguments are omitted to keep the notation concise.
Here $\bm{j}^r_{t}(\bm{r},\bm{v},\bm{y})$ is the $\bm{r}$-component of the probability current and similar for $\bm{v}$ and $\bm{y}$; the superscript S+F indicates the joint state of system (S) and feedback controller (F).
The two main assumption entering \eqref{continuity} are that (i) the joint dynamics of the system and the feedback controller are Markovian and that (ii) the noises acting on the system and feedback controller are independent---the latter assumption is referred to as the bipartite condition \cite{Che19}.
The $\bm{r}$ and $\bm{v}$ components of the probability current are explicitly given by \cite{Ris86} $\bm{j}^r_{t} = \bm{v} p^{\text{S} + \text{F}}_t$ and 
\begin{align}
m \bm{j}^v_{t} &= \bigg( - \grad_r U - \gamma \bm{v} + \bm{f}^\text{fb} - \frac{\gamma T}{m} \grad_v  \bigg) p^{\text{S} + \text{F}}_t .
\end{align}
For the $\bm{y}$-component of the probability current, no specific form beyond the bipartite condition is assumed.

\subsection*{Information thermodynamics for feedback cooling}
Using \eqref{continuity}, the overall dissipation of the system and the feedback controller can be characterized.
Assuming that the degrees of freedom of the feedback controller that enter the feedback force are even under time-reversal---the feedback controller may have odd internal degrees of freedom---the overall entropy production rate of the system is given by \cite{Hor14}
\begin{align}
\sigma_t^\text{S+F} = \sigma_t^\text{S} + \sigma_t^\text{F} \label{entropy-decomposition}
\end{align}
where the contribution from the system's degrees of freedom is
\begin{align}
\sigma_t^\text{S} = \frac{m^2}{\gamma T} \Av{\frac{\Vert \bm{j}_{t}^{v,\text{irr}} \Vert^2}{(p_t^{\text{S} + \text{F}})^2}}, \; \bm{j}_{t}^{v,\text{irr}} = - \frac{\gamma}{m} \bigg( \bm{v} + \frac{T}{m} \grad_v \bigg) p_t^{\text{S} + \text{F}} .
\end{align}
While the contribution from the feedback controller $\sigma_t^\text{F}$ depends on the precise form of the latter's dynamics, it is always positive.
Equivalently, the entropy production rate can be decomposed into the change in Shannon entropy and the entropy change of the environment,
\begin{align}
\sigma_t^\text{S+F} = \dot{S}_t^\text{S+F} + \sigma_t^\text{env,S+F} \geq 0, \label{entropy-environment}
\end{align}
where $S_t^\text{S+F} = - \av{\ln p_t^\text{S+F}}$ is the Gibbs-Shannon entropy of the joint probability density.
This is the second law of thermodynamics, identifying $\sigma_t^\text{env,S+F} = - \dot{Q}^\text{S+F}_t/T$ with the rate of rate of heat $\dot{Q}^\text{S+F}_t$ exchanged by both the system and feedback controller with the environment at temperature $T$.
The connection between \eqref{entropy-decomposition} and \eqref{entropy-environment} is provided by the second law of information thermodynamics \cite{Hor14}, 
\begin{align}
\sigma_t^\text{S} = \dot{S}_t^\text{S} + \sigma_t^\text{env,S} - l_t^\text{S} \geq 0,  \label{second-law-info}
\end{align}
and similar for the feedback controller, where the learning rate $l_t^\text{S}$ quantifies the change in mutual information $I_t^{\text{S}:\text{F}}$ due to the degrees of freedom of the system (see Appendix A),
\begin{align}
\dot{I}_t^\text{S:F} = l_t^\text{S} + l_t^\text{F} \label{learning-rates} .
\end{align}
In the presence of information flow, one subsystem can apparently violate the second law at the expense of the other subsystem---cooling the system S increases the dissipation of the feedback controller F so that the overall system satisfies the second law.
The environmental entropy change due to the system is
\begin{align}
\sigma_t^\text{env,S} = \frac{\gamma d}{m} \bigg( \frac{T_\text{K}}{T} - 1 \bigg) = - \frac{\dot{Q}_t^\text{S}}{T} \label{system-heat} ,
\end{align}
while the learning rate can be written as
\begin{align}
l_t^\text{S} = -\sigma_t^\text{epu} - l_t^\text{ex} \label{learning-rate-decomposition} ,
\end{align}
with the excess information flow \eqref{excess-definition} and entropy pumping rate \eqref{epu-definition}.
This allows writing \eqref{second-law-info} as
\begin{align}
\sigma_t^\text{S} = \dot{S}_t^\text{S} + \frac{\gamma d}{m} \bigg( \frac{T_\text{K}}{T} - 1 \bigg) + \sigma_t^\text{epu} + l_t^\text{ex} \geq 0 \label{second-law-excess-epu} .
\end{align}
In a similar manner it is found that (see Appendix B)
\begin{align}
\bar{\sigma}_t^\text{S} = \dot{S}_t^\text{S} + \frac{\gamma d}{m} \bigg( \frac{T_\text{K}}{T} - 1 \bigg) + \sigma_t^\text{epu} \geq 0 \label{second-law-epu} ,
\end{align}
where the effective entropy production rate is defined as
\begin{align}
\bar{\sigma}_t^\text{S} = \frac{m^2}{\gamma T} \Av{\frac{\Vert \bar{\bm{j}}_{t}^{v,\text{irr}} \Vert^2}{(p_t^{\text{S}})^2}}, \qquad \bar{\bm{j}}_{t}^{v,\text{irr}} = \int d\bm{y} \ \bm{j}_{t}^{v,\text{irr}} p^{\text{S}} \label{entropy-effective} .
\end{align}
Comparing \eqref{second-law-excess-epu} and \eqref{second-law-epu}, the second law holds with or without the excess information flow.
Thus, while $l_t^\text{ex}$ contributes to the information flow between the system and feedback controller, a negative heat flow due to cooling the system must be compensated solely by the entropy pumping rate.
Using \eqref{learning-rates}, the learning rate of the feedback controller is written as
\begin{align}
l_t^\text{F} = \dot{I}_t^{\text{S}:\text{F}} + \sigma_t^\text{epu} + l_t^\text{ex} .
\end{align}
In the steady state, the first term vanishes and \eqref{excess-epu-relation} is obtained.

\subsection*{Trade-off relations and bounds}
Applying the Cauchy-Schwarz inequality to \eqref{entropy-effective} yields
\begin{align}
\bar{\sigma}_t^\text{S} \geq \frac{m^2}{\gamma T} \frac{\Av{ \frac{\bm{u}  \cdot \bar{\bm{j}}_{t}^{v,\text{irr}}}{p_t^\text{S}}}^2}{\Av{\Vert \bm{u} \Vert^2}} \label{cauchy-schwarz}
\end{align}
for any vector field $\bm{u}(\bm{r},\bm{v})$.
Choosing $\bm{u}(\bm{r},\bm{v}) = \bm{v}$, this results in the lower bound on the effective entropy production rate
\begin{align}
\bar{\sigma}_t^\text{S} \geq \frac{\gamma d}{m} \bigg( \frac{T_\text{K}}{T} + \frac{T}{T_\text{K}} - 2 \bigg) \label{entropy-bound-effective} .
\end{align}
Since $\sigma_t^\text{S+F} \geq \sigma_t^\text{S} \geq \bar{\sigma}_t^\text{S}$, this provides a lower bound on the overall entropy production rate in terms of the kinetic temperature.
In particular, achieving a vanishing kinetic temperature necessarily requires a diverging dissipation rate.
Applying this to \eqref{second-law-epu} yields a lower bound on the kinetic temperature
\begin{align}
\frac{T_\text{K}}{T} \geq \frac{1}{1 + \frac{m}{\gamma d} \big( \sigma_t^\text{epu} + \dot{S}_t^\text{S} \big)} ,
\end{align}
which reduces to the inequality in \eqref{epu-temperature-bound} in the steady state.
The entropy pumping rate \eqref{epu-definition} can be written as
\begin{align}
\sigma_t^\text{epu} &= - \frac{1}{m} \Av{\grad_v  \cdot \bar{\bm{f}}_\text{fb}} \\
&= - \frac{1}{m} \Av{\big( \bm{f}_\text{fb} - \bar{\bm{f}}_\text{fb} \big)  \cdot \grad_v \ln p_t^{\text{F} \vert \text{S}}} \n,
\end{align}
using that $\av{\bm{u}  \cdot \grad_v \ln p_t^{\text{F} \vert \text{S}}} = 0$ for any vector field $\bm{u}(\bm{r},\bm{v})$ that does not depend on $\bm{y}$.
Applying the Cauchy-Schwarz inequality to the second expression yields \eqref{excess-epu-tradeoff}.

If the feedback controller is in contact with the same environment at temperature $T$, the second law \eqref{second-law-info} yields
\begin{align}
\frac{\dot{Q}^\text{F}_t}{T} = \sigma_t^\text{F} - \dot{S}_t^\text{F} + l_t^\text{F} = \sigma_t^\text{F} - \dot{S}_t^\text{F} + \dot{I}_t^{\text{S}:\text{F}} + \sigma_t^\text{epu} + l_t^\text{ex} .
\end{align}
Since $\sigma_t^\text{F}$ is positive, it holds that in the steady state
\begin{align}
\dot{Q}^\text{F} \geq T \big( \sigma^\text{epu} + l^\text{ex} \big),
\end{align}
where omitting the subscript $t$ implies the steady state.
Using \eqref{epu-temperature-bound}, this yields \eqref{dissipation-bound}.
The first law states
\begin{align}
\dot{E}_t^\text{S} = \dot{Q}_t^\text{S} + \dot{W}_t^\text{fb},
\end{align}
where $E_t^\text{S} = m \av{\Vert \bm{v} \Vert^2}/2 + \av{U}$ is the average energy and
\begin{align}
\dot{W}_t^\text{fb} = \Av{\bm{f}^\text{fb} \cdot \bm{v}} = \Av{\bar{\bm{f}}^\text{fb} \cdot \bm{v}}
\end{align}
is the work done by the feedback force on the particles.
In the steady state $\dot{E}_t^\text{S} = 0$ and therefore, using \eqref{second-law-epu}
\begin{align}
\Av{\bar{\bm{f}}^\text{fb} \cdot \bm{v}} = - \dot{Q}^\text{S} = T \big( \bar{\sigma}^\text{S} - \sigma^\text{epu} \big) \label{first-law}.
\end{align}
Now recall \eqref{cauchy-schwarz} and set $\bm{u} = \bar{\bm{f}}^\text{fb}$ to obtain
\begin{align}
\Av{\Vert \bar{\bm{f}}^\text{fb} \Vert^2} \geq \frac{\gamma \big( \av{\bar{\bm{f}}^\text{fb} \cdot \bm{v}} - \frac{T}{m} \av{\grad_v \cdot \bar{\bm{f}}^\text{fb}} \big)^2 }{T \bar{\sigma}^\text{S}} = \gamma T \bar{\sigma}^\text{S} \label{effective-feedback-force-bound} ,
\end{align}
where \eqref{epu-definition} was used.
We then have for the magnitude of the feedback force
\begin{align}
\Av{\Vert \bm{f}^\text{fb} \Vert^2} \geq \gamma T \bigg( \bar{\sigma}^\text{S} + \frac{(\sigma^\text{epu})^2}{l^\text{ex}} \bigg),
\end{align}
using $\av{\Vert \bm{f}^\text{fb} \Vert^2} = \av{\Vert \bar{\bm{f}}^\text{fb} \Vert^2} + \av{\Vert \delta \bm{f}^\text{fb} \Vert^2}$ and \eqref{excess-epu-tradeoff} to bound the second term.
Using the lower bound \eqref{entropy-bound-effective} for $\bar{\sigma}^\text{S}$ and \eqref{epu-temperature-bound} for $\sigma^\text{epu}$, this yields \eqref{force-bound}.

\subsection*{Optimality of linear feedback forces}
The solution of \eqref{langevin} in the presence of the feedback force is difficult to obtain, since it describes a non-linear system out of equilibrium.
A notable exception is when the effective feedback force is linear in the velocity $\bar{f}^\text{fb}(\bm{v}) = -\gamma^\text{fb} \bm{v}$.
Since this is of the same form as the environmental friction, the steady-state probability density of the system is a quasi-equilibrium distribution at temperature
\begin{align*}
T_\text{K} = \frac{\gamma}{\gamma + \gamma^\text{fb}} T .
\end{align*}
The entropy pumping rate in this case is $\sigma^\text{epu} = \gamma^\text{fb} d/m$, from which it follows that \eqref{epu-temperature-bound} reduces to an equality in this case, which corresponds to perfect cooling efficiency $\eta^\text{cool} = 1$.
Moreover, combining \eqref{entropy-bound-effective} and \eqref{effective-feedback-force-bound},
\begin{align}
\Av{\Vert \bar{\bm{f}}^\text{fb} \Vert^2} \geq \frac{\gamma^2 d T}{m} \bigg( \frac{T_\text{K}}{T} + \frac{T}{T_\text{K}} - 2 \bigg),
\end{align}
which provides a bound on the magnitude of the effective feedback force in terms of the kinetic temperature; equality is likewise attained for a linear feedback force.
In conclusion, a linear feedback force maximizes cooling for both a given entropy pumping rate and for a given overall magnitude of the force.

\subsection*{Scattering matrix and feedback force}
Assuming that the particle can be viewed as stationary from the point of view of the light field, the spatial dependence of the electric field $E(x)$ obeys the time-independent wave equation
\begin{align}
-\partial_x^2 E(x) - k^2 E(x) = 0,
\end{align}
where $k = \omega/c$ is the wave number with $c$ the speed of light and $\omega$ the frequency of the light field.
The above equation describes the propagation of light through vacuum; inside the particle, $k$ is replaced by $q = \sqrt{\epsilon_r} k$, where $\epsilon_r$ is the relative permittivity of the material of the particle.
The overall light field then obeys
\begin{gather}
-\partial_x^2 E(x) + V(x) E(x) - k^2 E(x) = 0 \quad \text{with} \\
V(x) = k^2(1-\epsilon_r) \Theta \bigg( \frac{L}{2} - |x-r| \bigg) \n.
\end{gather}
Here, $r$ is the position of the particle, $L$ is its diameter and $\Theta(x)$ is the Heaviside-theta function.
This equation is equivalent to the scattering by a square well or barrier in quantum mechanics.
The scattering matrix is given by
\begin{align}
\bm{S}(r) = &\frac{e^{-i k L}}{\cos(q L) - i \sin(q L) \frac{q^2 + k^2}{2 q k}} \\
&\times \begin{pmatrix} i \sin(q L) e^{2 i k r} \frac{q^2 - k^2}{2 q k} & 1 \\ 1 & i \sin(q L) e^{-2 i k r} \frac{q^2 - k^2}{2 q k} \end{pmatrix} . \n 
\end{align}
Plugging this into \eqref{optical-force}, one finds
\begin{align}
&f^\text{fb} = \frac{k}{\omega} \frac{(1 - \epsilon_r) \sin(q L)}{4 \epsilon_r + (1 - \epsilon_r)^2 \sin(q L)^2}  \\
&\hspace{1cm} \times \bigg( \big( \vert c_{l,\text{in}} \vert^2 - \vert c_{r,\text{in}} \vert^2 \big) (1 - \epsilon_r) \sin(q L) \nn
& \hspace{2cm} + 2 \sqrt{\epsilon_r} \Im \big( c_{l,\text{in}} c_{r,\text{in}}^* e^{2 i k r(t)} \big) \bigg) \n ,
\end{align}
where $\Im$ denotes the imaginary part.
Parameterizing the incoming light amplitudes as in \eqref{light-modes}, this results in \eqref{feedback-force}.

\subsection*{Parameter values}
For Figs.~\ref{fig-temperature}, \ref{fig-temperature-noise} and \ref{fig-power}, the parameter values of Ref.~\cite{Li11} are used.
Specifically, the particle is a silica (density $\rho = \num{2.5e3} \si{kg/m^3}$, refractive index $n = \sqrt{\epsilon_r} = 1.46$) particle of diameter $L = 3.0 \si{\mu\metre}$, leading to a mass of $m = \num{3.2e-14}\si{kg}$.
The cooling laser has a wavelength of $\lambda = 2 \pi/k = \num{532} \si{nm}$, and the particle is trapped in a parabolic potential with trapping frequency $\Omega = \sqrt{\kappa/m} = 2 \pi \times \num{9.1} \si{kHz}$.
The environmental temperature is $T = \num{300} \si{K}$, and the environmental damping rate is $\Gamma/(2 \pi) = \gamma/(2 \pi m) = \num{2.0e2} \si{Hz}$ (corresponding to a pressure of $637 \si{Pa}$).
The correlation time of the measurement noise is set to $\tau^\text{noise} = 10^{-4} \tau^\text{osc}$, where $\tau^\text{osc} = 2\pi/\Omega$ is the period of the oscillations in the trap, which is the shortest timescale in the system.

\textit{Acknowledgements.} A.D. was supported by JSPS Kakenhi (Grants No. 22K13974, 24H00833 and 25K00926) and JST ERATO Grant Number JPMJER2302. S.I. is supported by JSPS KAKENHI (Grants No.\ 22H01141, 23H00467, and 24H00834), JST ERATO Grant No.\ JPMJER2302, and UTEC-UTokyo FSI Research Grant Program. J.H. and S.R. thank N. Bachelard for helpful discussions.

\bibliography{bib}

\appendix
\begin{widetext}

\section{Information thermodynamics for underdamped systems}

\subsection{Dynamics of system and feedback controller}
While the information thermodynamics for overdamped and Markov jump systems have been extensively discussed in the literature \cite{All09,Sag12,Ito13,Har14,Hor14,Par15}, underdamped systems have not been investigated to such an extent.
A notable exception is Ref.~\cite{Hor14b}, where various measures of information exchange were computed and compared for a simple linear system.
Here, we provide a self-contained derivation of the information thermodynamics of general systems and feedback processes.
We consider a system of $n$ particles with masses $(m_1,\ldots,m_n)$ in $3$ dimensions, comprising $d = 3n$ positions and velocities $\bm{r} = (r_{1,1}, r_{1,2}, r_{1,3}, \ldots, r_{n,1}, r_{n,2}, r_{n,3}) = (r_1, \ldots, r_d)$ and $\bm{v} = (v_1,\ldots,v_d)$.
The particles are in contact with a thermal environment at temperature $T$, with which they interact via Stokes friction with (possibly different) friction coefficients $(\gamma_1,\ldots,\gamma_n)$ and thermal noise.
A potential $U(\bm{r})$ describes (conservative) systematic external forces and interactions between the particles.
In addition, the particles are affected by a feedback force $\bm{f}^\text{fb}(\bm{r},\bm{y})$, which may depend on the positions (but not velocities) of the particles and a set of $q$ variables $\bm{y} = (y_1,\ldots,y_q)$, which we take to represent the current configuration of a feedback controller, that is, another physical system that determines the value of the feedback force applied to the system.
The equations of motion for the particles are the underdampled Langevin equations,
\begin{align}
\dot{\bm{r}}(t) = \bm{v}(t), \qquad \bm{m} \dot{\bm{v}}(t) = - \grad_r U(\bm{r}(t)) - \bm{\gamma} \bm{v}(t) + \bm{f}^\text{fb}(\bm{r}(t),\bm{y}(t)) + \sqrt{2 \bm{\gamma} T} \bm{\xi}(t) \label{langevin-sup} .
\end{align}
Here, $\bm{m}$ is a diagonal matrix containing the masses of the particles, $\bm{\gamma}$ is a diagonal matrix containing the friction coefficients and $\bm{\xi}(t)$ is a vector of mutually independent Gaussian white noises.
Note that since $\bm{m}$ and $\bm{\gamma}$ are both diagonal, they commute.
For concreteness, we assume that the configuration of the feedback controller evolves according to a similar but more general set of equations
\begin{align}
\dot{\bm{y}}(t) = \bm{g}(\bm{r}(t),\bm{v}(t),\bm{y}(t)) + \sqrt{2 \bm{D}(\bm{r}(t),\bm{v}(t),\bm{y}(t))} \bullet \bm{\xi}_y(t) \label{langevin-feedback}.
\end{align}
$\bm{g}(\bm{r},\bm{v},\bm{y})$ is a set of functions that can depend on the configurations of both the particles and feedback controller.
Likewise, the positive semi-definite matrix $\bm{D}(\bm{r},\bm{v},\bm{y})$, which determines the coupling to the Gaussian white noises $\bm{\xi}_y(t)$, may depend on the configuration of both the particles and feedback controller.
The matrix $\sqrt{\bm{D}(\bm{r},\bm{v},\bm{y})}$ is the unique positive semi-definite square root of $\bm{D}(\bm{r},\bm{v},\bm{y})$ and $\bullet$ denotes the Ito-product.
We remark that we may also include discrete degrees of freedom in the feedback controller without impacting the validity of the following results.
The central assumptions entering the above description are that the joint dynamics of the particles and feedback controller are Markovian and that the noises acting on the particles and feedback controller are independent.
However, in principle, \eqref{langevin-feedback} can also account for many types of effectively non-Markovian feedback, which can be represented as a Markovian dynamics in an extended state space by a technique known as Markovian embedding \cite{Sie10,Loo19,Kan20,Kan24}.
Equivalently to the stochastic description \eqref{langevin-sup} and \eqref{langevin-feedback}, we can describe the evolution of the joint probability density of the particle system (S) and feedback controller (F) $p_t^{\text{S}+\text{F}}(\bm{r},\bm{v},\bm{y})$, which follows the Fokker-Planck equation
\begin{align}
\partial_t p_t^{\text{S}+\text{F}}(\bm{r},\bm{v},\bm{y}) = - \grad_r \cdot \bm{j}_t^r(\bm{r},\bm{v},\bm{y}) - \grad_v  \cdot \bm{j}_t^v(\bm{r},\bm{v},\bm{y}) - \grad_y  \cdot \bm{j}_t^y(\bm{r},\bm{v},\bm{y}) \label{fpe},
\end{align}
where the probability currents are given by
\begin{subequations}
\begin{align}
\bm{j}_t^r(\bm{r},\bm{v},\bm{y}) &= \bm{v} p_t^{\text{S}+\text{F}}(\bm{r},\bm{v},\bm{y}), \\
\bm{j}_t^v(\bm{r},\bm{v},\bm{y}) &= \bm{m}^{-1} \Big( \big( - \grad_r U(\bm{r}) - \bm{\gamma} \bm{v} + \bm{f}^\text{fb}(\bm{r},\bm{y}) \big) p_t^{\text{S}+\text{F}}(\bm{r},\bm{v},\bm{y}) - \bm{m}^{-1} \bm{\gamma} T \grad_v p_t^{\text{S}+\text{F}}(\bm{r},\bm{v},\bm{y}) \Big), \\
\bm{j}_t^y(\bm{r},\bm{v},\bm{y}) &= \big( \bm{g}(\bm{r},\bm{v},\bm{y}) + \bm{g}_\text{D}(\bm{r},\bm{v},\bm{y}) \big) p_t^{\text{S}+\text{F}}(\bm{r},\bm{v},\bm{y}) - \bm{D}(\bm{r},\bm{v},\bm{y}) \grad_y p_t^{\text{S}+\text{F}}(\bm{r},\bm{v},\bm{y}) \\
&\text{with} \qquad  \big(\bm{g}_\text{D}(\bm{r},\bm{v},\bm{y}) \big)_{j} = - \sum_{k = 1}^q \partial_{y_j} D_{jk}(\bm{r},\bm{v},\bm{y}) \n  .
\end{align} \label{fpe-currents}%
\end{subequations}
Introducing the marginal probability of the system
\begin{align}
p_t^\text{S}(\bm{r},\bm{v}) = \int d\bm{y} \ p_t^{\text{S}+\text{F}}(\bm{r},\bm{v},\bm{y})
\end{align}
and integrating \eqref{fpe} with respect to the degrees of freedom of the feedback controller, we further obtain
\begin{align}
\partial_t p_t^{\text{S}}(\bm{r},\bm{v}) = - \grad_r  \cdot \bar{\bm{j}}_t^r(\bm{r},\bm{v}) - \grad_v  \cdot \bar{\bm{j}}_t^v(\bm{r},\bm{v}) \label{fpe-system},
\end{align}
where the effective probability currents are given by
\begin{subequations}
\begin{align}
\bar{\bm{j}}_t^r(\bm{r},\bm{v}) &= \bm{v} p_t^{\text{S}}(\bm{r},\bm{v}), \\
\bar{\bm{j}}_t^v(\bm{r},\bm{v}) &= \bm{m}^{-1} \Big( \big( - \grad_r U(\bm{r}) - \bm{\gamma} \bm{v} + \bar{\bm{f}}^\text{fb}(\bm{r},\bm{v}) \big) p_t^{\text{S}}(\bm{r},\bm{v}) - \bm{m}^{-1} \bm{\gamma} T \grad_v p_t^{\text{S}}(\bm{r},\bm{v}) \Big) .
\end{align} \label{fpe-currents-system}%
\end{subequations}
The effective feedback force is given by
\begin{align}
\bar{\bm{f}}_t^\text{fb}(\bm{r},\bm{v}) = \int d\bm{y} \ \bm{f}^\text{fb}(\bm{r},\bm{y}) p_t^{\text{F} \vert \text{S}}(\bm{y} \vert \bm{r},\bm{v}) \qquad \text{with} \qquad p_t^{\text{F} \vert \text{S}}(\bm{y} \vert \bm{r},\bm{v}) = \frac{p_t^{\text{S}+\text{F}}(\bm{r},\bm{v},\bm{y})}{p_t^{\text{S}}(\bm{r},\bm{v})} .
\end{align}
Note that while the actual feedback force is independent of the system's velocity, the effective feedback force may depend on velocity via the velocity-dependence of the feedback controller's state.
Likewise, even though the actual feedback force does not explicitly depend on time, the effective feedback force may be time-dependent via the dynamics.
While \eqref{fpe-system} is apparently an equation only involving the degrees of freedom of the system, it implicitly depends on the solution of the full dynamics \eqref{fpe} via the conditional probability density of the feedback controller $p_t^{\text{F} \vert \text{S}}(\bm{y} \vert \bm{r},\bm{v})$.
This represents the fact that the dynamics of the system by itself (without knowledge about the feedback controller) is non-Markovian.
As a consequence, while \eqref{fpe-system} is formally equivalent to the Langevin equation
\begin{align}
\dot{\bm{r}}(t) = \bm{v}(t), \qquad \bm{m} \dot{\bm{v}}(t) = - \grad_r U(\bm{r}(t)) - \bm{\gamma} \bm{v}(t) + \bar{\bm{f}}_t^\text{fb}(\bm{r}(t),\bm{v}(t)) + \sqrt{2 \bm{\gamma} T} \bm{\xi}(t) \label{langevin-system},
\end{align}
this equivalence only holds on the level of the probability density $p_t^{\text{S}}(\bm{r},\bm{v})$ and does not extend to correlations or higher-order statistics.
However, as long as we are interested in statistics that can be expressed in terms of $p_t^{\text{S}}(\bm{r},\bm{v})$, \eqref{langevin-system} shows that we can describe the effect of the feedback as an effective, velocity-dependent force.

\subsection{Entropy production and thermodynamics}
We write the entropy production rate corresponding to the dynamics \eqref{fpe} as \cite{Spi12b,Spi12}
\begin{align}
\sigma_t^\text{S+F} &= \sigma_t^\text{S} + \sigma_t^\text{F} \qquad \text{with} \qquad
\sigma_t^\text{S} = \frac{1}{T} \Av{\bigg(\frac{\bm{j}_t^{v,\text{irr}}}{p_t^{\text{S}+\text{F}}} \bigg) \cdot \bm{m}^2 \bm{\gamma}^{-1} \bigg(\frac{\bm{j}_t^{v,\text{irr}}}{p_t^{\text{S}+\text{F}}} \bigg)} \label{entropy} .
\end{align}
Here, we introduced the irreversible part of the velocity probability current
\begin{align}
\bm{j}_t^{v,\text{irr}}(\bm{r},\bm{v},\bm{y}) &= \bm{m}^{-1} \Big( - \bm{\gamma} \bm{v} p_t^{\text{S}+\text{F}}(\bm{r},\bm{v},\bm{y}) - \bm{m}^{-1} \bm{\gamma} T \grad_v p_t^{\text{S}+\text{F}}(\bm{r},\bm{v},\bm{y}) \Big).
\end{align}
From the explicit expression, we see that $\sigma_t^\text{S} \geq 0$.
Likewise, $\sigma_t^\text{F} \geq 0$, however, its concrete expression depends on the precise dynamics \eqref{langevin-feedback}, in particular, on whether the latter contains odd (velocity-like) degrees of freedom under time-reversal or not.
If all degrees of freedom of the feedback controller are even under time-reversal, then \eqref{langevin-feedback} corresponds to overdamped dynamics and
\begin{align}
\sigma_t^\text{F} = \Av{ \bigg( \frac{\bm{j}_t^y}{p_t^{\text{S}+\text{F}}} \bigg) \cdot \bm{D}^{-1} \bigg( \frac{\bm{j}_t^y}{p_t^{\text{S}+\text{F}}} \bigg)} .
\end{align}
If the dynamics of the feedback controller contain odd degrees of freedom, then we have to identify the irreversible part of the probability currents as described in Refs.~\cite{Spi12b,Spi12}.
We can also write \eqref{entropy} as as a formal second law of thermodynamics
\begin{align}
\sigma_t^\text{S+F} = \dot{S}_t^\text{S+F} + \sigma_t^\text{env,S+F}, \label{second-law}
\end{align}
where the dot denotes differentiation with respect to time.
The Gibbs-Shannon entropy of the system and feedback controller is defined 
\begin{align}
S^{\text{S}+\text{F}}_t = - \Av{\ln p_t^{\text{S}+\text{F}}}
\end{align}
and $\sigma_t^\text{env,S+F}$ is a \enquote{heat-like} contribution that describes the change in the entropy of the environment.
However, the identification of this term with the heat exchanged between the system and environment is only possible if the noise acting on the system stems from a thermal environment, which is not necessarily the case for the feedback controller.

Applying the same argument to \eqref{fpe-system}, we define the entropy production rate of the effective dynamics \eqref{langevin-system}
\begin{align}
\bar{\sigma}_t^\text{S} &= \frac{1}{T} \Av{\bigg(\frac{\bar{\bm{j}}_t^{v,\text{irr}}}{p_t^{\text{S}}} \bigg) \cdot \bm{m}^2 \bm{\gamma}^{-1} \bigg(\frac{\bar{\bm{j}}_t^{v,\text{irr}}}{p_t^{\text{S}}} \bigg)} \label{entropy-system} \qquad \text{with} \\
\bar{\bm{j}}_t^{v,\text{irr}}(\bm{r},\bm{v}) &= \int d\bm{y} \ \bm{j}_t^{v,\text{irr}}(\bm{r},\bm{v},\bm{y}) = \bm{m}^{-1} \Big(  - \bm{\gamma} \bm{v} p_t^{\text{S}}(\bm{r},\bm{v}) - \bm{m}^{-1} \bm{\gamma} T \grad_v p_t^{\text{S}}(\bm{r},\bm{v}) \Big) \n .
\end{align}
We now derive two equivalent expressions for $\bar{\sigma}_t^\text{S}$.
In the following we assume either natural boundary conditions with respect to $\bm{r}$ and $\bm{v}$ ($p_t(\bm{r},\bm{v})$ vanishes sufficiently fast as $\Vert \bm{r} \Vert \rightarrow \infty$ and $\Vert \bm{v} \Vert \rightarrow \infty$) or natural boundary conditions for $\bm{v}$ and periodic boundary conditions for $\bm{r}$.
Doing so will allow us to integrate by parts with respect to $\bm{r}$ and $\bm{v}$ without considering boundary terms.
First, using the explicit expression of $\bar{\bm{j}}_t^{v,\text{irr}}(\bm{r},\bm{v})$ and expanding the square, we obtain
\begin{align}
\bar{\sigma}_t^\text{S} = \frac{1}{T} \Av{\bm{v} \cdot \bm{\gamma} \bm{v}} + T \Av{\grad_v \ln p_t^\text{S} \bm{\gamma} \bm{m}^{-2} \grad_v \ln p_t^\text{S}} + 2 \int d\bm{r} \int d\bm{v} \ \bm{v} \cdot \bm{\gamma} \bm{m}^{-1} \grad_v p_t^\text{S}(\bm{r},\bm{v}) .
\end{align}
Integrating by parts in the last term, we find
\begin{align}
\bar{\sigma}_t^\text{S} = \frac{1}{T} \Av{\bm{v} \cdot \bm{\gamma} \bm{v}} + T \Av{\grad_v \ln p_t^\text{S} \bm{\gamma} \bm{m}^{-2} \grad_v \ln p_t^\text{S}} - 2 \text{tr}\big( \bm{\gamma} \bm{m}^{-1} \big) \label{entropy-system-1} ,
\end{align}
where tr denotes the trace of a matrix.
On the other hand, only expanding one instance of $\bar{\bm{j}}_t^{v,\text{irr}}(\bm{r},\bm{v})$, we can express \eqref{entropy-system} as
\begin{align}
\bar{\sigma}_t^\text{S} &= - \int d\bm{r} \int d\bm{v} \ \bigg( \frac{1}{T} \big(\bm{m} \bm{v} \big) \cdot \bar{\bm{j}}_t^{v,\text{irr}}(\bm{r},\bm{v}) + \grad_v \ln p_t^\text{S}(\bm{r},\bm{v})  \cdot \bar{\bm{j}}_t^{v,\text{irr}}(\bm{r},\bm{v}) \bigg) .
\end{align}
We plug in the expression for $\bar{\bm{j}}_t^{v,\text{irr}}(\bm{r},\bm{v})$ in the first term and integrate by parts in the second term,
\begin{align}
\bar{\sigma}_t^\text{S} &= \int d\bm{r} \int d\bm{v} \ \bigg( \frac{1}{T} \bm{v} \cdot \Big( \bm{\gamma} \bm{v} p_t^{\text{S}}(\bm{r},\bm{v}) + \bm{m}^{-1} \bm{\gamma} T \grad_v p_t^{\text{S}}(\bm{r},\bm{v}) \Big) + \ln p_t^\text{S}(\bm{r},\bm{v}) \grad_v \cdot \bar{\bm{j}}_t^{v,\text{irr}}(\bm{r},\bm{v}) \bigg) .
\end{align}
We now use \eqref{fpe-system} to replace the divergence of the irreversible velocity current,
\begin{align}
\bar{\sigma}_t^\text{S} = \int d\bm{r} \int d\bm{v} \ \bigg( &\frac{1}{T} \bm{v} \cdot \Big( \bm{\gamma} \bm{v} p_t^{\text{S}}(\bm{r},\bm{v}) + \bm{m}^{-1} \bm{\gamma} T \grad_v p_t^{\text{S}}(\bm{r},\bm{v}) \Big) \\
& + \ln p_t^\text{S}(\bm{r},\bm{v}) \Big( - \partial_t p_t^{S}(\bm{r},\bm{v}) -  \grad_r  \cdot \bm{v} p_t^{S}(\bm{r},\bm{v}) - \grad_v \cdot \bm{m}^{-1} \Big( \big(- \grad_r U(\bm{r}) + \bar{\bm{f}}_t^\text{fb}(\bm{r},\bm{v}) \big) p_t^{S}(\bm{r},\bm{v}) \Big)  \bigg) .  \n
\end{align}
We identify the change in the Shannon entropy of the system
\begin{align}
- \int d\bm{r} \int d\bm{v} \ \ln p_t^\text{S}(\bm{r},\bm{v}) \partial_t p_t^{S}(\bm{r},\bm{v}) = - d_t \Av{\ln p_t^\text{S}} = \dot{S}_t^\text{S} .
\end{align}
Integrating by parts once more in the second and third term of the second line, we obtain
\begin{align}
\bar{\sigma}_t^\text{S} = \dot{S}_t^\text{S} + \int d\bm{r} \int d\bm{v} \ \bigg( &\frac{1}{T} \bm{v} \cdot \Big( \bm{\gamma} \bm{v} p_t^{\text{S}}(\bm{r},\bm{v}) + \bm{m}^{-1} \bm{\gamma} T \grad_v p_t^{\text{S}}(\bm{r},\bm{v}) \Big)  \\
& + \Big(  \bm{v} \cdot \grad_r  \ln p_t^\text{S}(\bm{r},\bm{v}) +  \bm{m}^{-1} \big(- \grad_r U(\bm{r}) + \bar{\bm{f}}_t^\text{fb}(\bm{r},\bm{v}) \big) \cdot \grad_v \ln p_t^\text{S}(\bm{r},\bm{v}) \Big) p_t^\text{S}(\bm{r},\bm{v}) \bigg) .  \n
\end{align}
The first term and the term involving the potential in the second line vanish due to the boundary conditions.
Finally, we integrate by parts in the second term of the first line and in the remaining term in the second line to obtain
\begin{align}
\bar{\sigma}_t^\text{S} = \dot{S}_t^\text{S} + \sigma_t^\text{epu} + \frac{1}{T} \Big( \Av{\bm{v} \cdot \bm{\gamma} \bm{v}} - T \text{tr}\big( \bm{\gamma} \bm{m}^{-1} \big) \Big)  \label{entropy-system-2} .
\end{align}
Here, tr denotes the trace of a matrix and we defined the entropy pumping rate
\begin{align}
\sigma_t^\text{epu} = - \Av{\grad_v \cdot \bm{m}^{-1} \bar{\bm{f}}_t^\text{fb}} \label{entropy-pumping} .
\end{align}
The entropy pumping rate explicitly quantifies the velocity-dependence of the effective feedback force, it vanishes whenever $\bar{f}^\text{fb}(\bm{r},\bm{v})$ is independent of $\bm{v}$.
We identify the second term in \eqref{entropy-system-2} as the change in the environment due to the dynamics of the system,
\begin{align}
\sigma_t^\text{env,S} = \frac{1}{T} \Big( \Av{\bm{v} \cdot \bm{\gamma} \bm{v}} - T \text{tr}\big( \bm{\gamma} \bm{m}^{-1} \big) \Big) = - \frac{1}{T} \mathcal{Q}_t^\text{S} ,
\end{align}
where $\mathcal{Q}_t^\text{S}$ denotes the rate of heat transfer from the thermal environment to the system, which is equal to the rate of work done on the system by the friction and noise forces \cite{Sek10,Sei12},
\begin{align}
\mathcal{Q}_t^\text{S} = \Av{\big(-\bm{\gamma} \bm{v} + \sqrt{2 \bm{\gamma T}} \bm{\xi} \big) \circ \bm{v}}, 
\end{align}
where $\circ$ denotes the Stratonovich product.
Using \eqref{langevin-sup}, this is equivalent to
\begin{align}
\mathcal{Q}_t^\text{S} =  \Av{ \big( \bm{m} \dot{\bm{v}} + \grad_r U - \bm{f}^\text{fb} \big) \circ \bm{v}} = d_t \bigg( \frac{1}{2} \Av{\bm{v} \cdot \bm{m} \bm{v}} + \Av{U} \bigg) - \Av{\bm{f}^\text{fb} \cdot \bm{v}} .
\end{align}
Identifying the sum of kinetic and potential energy as the total energy $E_t^\text{S}$ of the system, this yields the first law of thermodynamics,
\begin{align}
\dot{E}_t^\text{S} = \mathcal{Q}_t^\text{S} + \mathcal{W}_t^\text{fb},
\end{align}
where $\mathcal{W}_t^\text{fb} = \av{\bm{f}^\text{fb} \cdot \bm{v}} = \av{\bar{\bm{f}}_t^\text{fb} \cdot \bm{v}}$ is the rate of work done by the feedback force.
We note for later use that we can also rewrite the change in the system's Shannon entropy using the entropy pumping rate,
\begin{align}
\dot{S}_t^\text{S} + \sigma_t^\text{epu} = T \Av{\grad_v \ln p_t^\text{S} \cdot \bm{\gamma} \bm{m}^{-2} \grad_v \ln p_t^\text{S}} - T \text{tr}\big( \bm{\gamma} \bm{m}^{-1} \big),
\end{align}
which follows by comparing \eqref{entropy-system-1} and \eqref{entropy-system-2}.

We now relate the entropy production rates in the joint dynamics of the system and feedback controller, \eqref{entropy}, and of the effective dynamics of just the system, \eqref{entropy-system}.
For this, we note the relation from the definition \eqref{entropy-system}
\begin{align}
\frac{\bar{\bm{j}}_t^{v,\text{irr}}(\bm{r},\bm{v})}{p_t^\text{S}(\bm{r},\bm{v})} = \int d\bm{y} \ \frac{\bm{j}_t^{v,\text{irr}}(\bm{r},\bm{v},\bm{y})}{p_t^\text{S+F}(\bm{r},\bm{v},\bm{y})} p_t^{\text{F} \vert \text{S}}(\bm{y} \vert \bm{r},\bm{v}) .
\end{align}
In other words, the expression on the left-hand side is the conditional average of $\bm{j}_t^{v,\text{irr}}(\bm{r},\bm{v},\bm{y})/p_t^\text{S+F}(\bm{r},\bm{v},\bm{y})$ with respect to $\bm{y}$.
This allows us to write
\begin{align}
\sigma_t^\text{S} &= \frac{1}{T} \Av{\bigg(\frac{\bm{j}_t^{v,\text{irr}}}{p_t^{\text{S}+\text{F}}} \bigg) \cdot \bm{m}^2 \bm{\gamma}^{-1} \bigg(\frac{\bm{j}_t^{v,\text{irr}}}{p_t^{\text{S}+\text{F}}} \bigg)} \\
& = \frac{1}{T} \Av{\bigg(\frac{\bm{j}_t^{v,\text{irr}}}{p_t^{\text{S}+\text{F}}} - \frac{\bar{\bm{j}}_t^{v,\text{irr}}}{p_t^\text{S}} \bigg) \cdot \bm{m}^2 \bm{\gamma}^{-1} \bigg(\frac{\bm{j}_t^{v,\text{irr}}}{p_t^{\text{S}+\text{F}}} - \frac{\bar{\bm{j}}_t^{v,\text{irr}}}{p_t^\text{S}} \bigg)} + \frac{1}{T} \Av{\bigg(\frac{\bar{\bm{j}}_t^{v,\text{irr}}}{p_t^{\text{S}}} \bigg) \cdot \bm{m}^2 \bm{\gamma}^{-1} \bigg(\frac{\bar{\bm{j}}_t^{v,\text{irr}}}{p_t^{\text{S}}} \bigg)} \nn
&= \bar{\sigma}_t^\text{S} + \frac{1}{T} \Av{\bigg(\frac{\bm{j}_t^{v,\text{irr}}}{p_t^{\text{S}+\text{F}}} - \frac{\bar{\bm{j}}_t^{v,\text{irr}}}{p_t^\text{S}} \bigg) \cdot \bm{m}^2 \bm{\gamma}^{-1} \bigg(\frac{\bm{j}_t^{v,\text{irr}}}{p_t^{\text{S}+\text{F}}} - \frac{\bar{\bm{j}}_t^{v,\text{irr}}}{p_t^\text{S}} \bigg)} \n .
\end{align}
Since the second term is positive, we therefore have $\sigma_t^\text{S} \geq \bar{\sigma}_t^\text{S}$.
We further find for the difference between the two
\begin{align}
\sigma_t^\text{S} - \bar{\sigma}_t^\text{S} = \frac{1}{T} \Av{\bigg(\frac{\bm{j}_t^{v,\text{irr}}}{p_t^{\text{S}+\text{F}}} - \frac{\bar{\bm{j}}_t^{v,\text{irr}}}{p_t^\text{S}} \bigg) \cdot \bm{m}^2 \bm{\gamma}^{-1} \bigg(\frac{\bm{j}_t^{v,\text{irr}}}{p_t^{\text{S}+\text{F}}} - \frac{\bar{\bm{j}}_t^{v,\text{irr}}}{p_t^\text{S}} \bigg)} = T \Av{\grad_v \ln p_t^{\text{F} \vert \text{S}} \bm{\gamma} \bm{m}^{-2} \grad_v \ln p_t^{\text{F} \vert \text{S}}} .
\end{align}
Using this and \eqref{entropy-system-2}, we can write
\begin{align}
\sigma_t^\text{S} = \underbrace{\dot{S}_t^\text{S} - \frac{\dot{Q}_t^\text{S}}{T} + \sigma_t^\text{epu}}_{= \bar{\sigma}_t^\text{S}} + T \Av{\grad_v \ln p_t^{\text{F} \vert \text{S}} \bm{\gamma} \bm{m}^{-2} \grad_v \ln p_t^{\text{F} \vert \text{S}}} . \label{entropy-2}
\end{align}
In other words, the entropy production rate of the system in the joint dynamics \eqref{fpe} can be decomposed into the change of Shannon entropy of the system, the heat exchanged between the system and the environment, the entropy pumping rate and an additional, positive term that quantifies the velocity-dependence of the state of the feedback controller conditioned on the system.
The first three contributions correspond to the (likewise positive) entropy production rate of the system's effective dynamics.

\subsection{Fisher information}
We can express the above relations by using the Fisher information matrix with respect to $\bm{v}$.
We define
\begin{subequations}
\begin{align}
\big(\bmc{F}^{\text{S}+\text{F}}_{t,v} \big)_{jk} &= \Av{\partial_{v_j} \ln p_t^{\text{S}+\text{F}} \partial_{v_k} \ln p_t^{\text{S}+\text{F}}}, \\
\big(\bmc{F}^{\text{S}}_{t,v} \big)_{jk} &= \Av{\partial_{v_j} \ln p_t^{\text{S}} \partial_{v_k} \ln p_t^{\text{S}}}, \\
\big(\bmc{F}^{\text{F} \vert \text{S}}_{t,v} \big)_{jk} &= \Av{\partial_{v_j} \ln p_t^{\text{F} \vert \text{S}} \partial_{v_k} \ln p_t^{\text{F} \vert \text{S}}} .
\end{align}\label{fisher}%
\end{subequations}
$\bmc{F}_{t,v}^{\text{S}+\text{F}}$ is the Fisher information matrix of the joint probability density with respect to velocity, $\bmc{F}^{\text{S}}_{t,v}$ is the Fisher information matrix of the system and $\bmc{F}^{\text{F} \vert \text{S}_{t,v}}$ is the Fisher information matrix of the feedback controller conditioned on the system. 
These are positive definite matrices that quantify the dependence of the respective probability density on velocity; in other words, the information about the system's velocities encoded in the respective probability density.
They further satisfy the chain rule of the Fisher information
\begin{align}
\bmc{F}^{\text{S}+\text{F}}_{t,v} = \bmc{F}^{\text{F} \vert \text{S}}_{t,v} + \bmc{F}^{\text{S}}_{t,v} \label{fisher-chain-rule-sup} .
\end{align}
Using this, we can rewrite \eqref{entropy-system-1} as
\begin{align}
\bar{\sigma}_t^\text{S} = \text{tr} \Bigg( \bm{\gamma} \bm{m}^{-1} \bigg( \frac{1}{T} \bm{m} \bmc{V}_t + T \bm{m}^{-1} \bmc{F}_{t,v}^\text{S} - 2 \bm{I} \bigg) \Bigg) ,
\end{align}
where $\bmc{V}_t$ is the matrix of second moments of the velocity,
\begin{align}
\big(\bmc{V}_t \big)_{jk} = \Av{v_j v_k} \label{velocity-covariance} ,
\end{align}
and $\bm{I}$ denotes the identity matrix.
The change in Shannon entropy can likewise be expressed as
\begin{align}
\dot{S}_t^\text{S} + \sigma_t^\text{epu} = \text{tr} \Big( \bm{\gamma} \bm{m}^{-1} \big( T \bm{m}^{-1} \bmc{F}_{t,v}^{\text{S}} - \bm{I} \big) \Big) \label{shannon-fisher} .
\end{align}
This relates the change in Shannon entropy and the entropy pumping rate to the Fisher information matrix of the system's state with respect to velocity.
The difference between the entropy production in the full and effective dynamics is
\begin{align}
\sigma_t^\text{S} - \bar{\sigma}_t^\text{S} = T \text{tr} \big( \bm{\gamma} \bm{m}^{-2} \bmc{F}_{t,v}^{\text{F} \vert \text{S}} \big) \label{fisher-entropy-difference} .
\end{align}
This expression, together with \eqref{fisher-chain-rule-sup} provides an information-theoretic interpretation of the difference between the entropy production rate in the two different levels of description of the system's dynamics.
The Fisher information $\bmc{F}_{t,v}^{\text{F} \vert \text{S}}$ quantifies the additional information about velocity contained in the state of the feedback controller, that is not already contained in the state of the system.
\eqref{fisher-entropy-difference} states that this additional information results in additional irreversibility of the joint dynamics, leading to increased dissipation.
We remark that this additional contribution to the entropy production rate is independent of the precise realization of the dynamics of the feedback controller, but only involves the resulting velocity-dependence of the latter.
Moreover, the additional contribution also appears in the absence of any feedback force; even if the feedback controller does not act on the system, the mere act of acquiring information about the system's velocity will lead to irreversibility of the overall dynamics.

This also allows us to treat the case where not all degrees of freedom of the feedback controller are resolved in \eqref{langevin-feedback}.
For example, assume the physical feedback controller consists of a part F with degrees of freedom $\bm{y}$, which determine the feedback force acting on the system, and an additional part $\tilde{\text{F}}$ representing some internal degrees of freedom $\bm{z}$ of the feedback controller whose dynamics are not known explicitly.
Then, we have using the chain rule of the Fisher information
\begin{align}
\bmc{F}^{\text{S}+\text{F}+\tilde{\text{F}}}_{t,v} = \bmc{F}^{\tilde{\text{F}} \vert \text{S}+\text{F}}_{t,v} + \bmc{F}^{\text{S}+\text{F}}_{t,v} = \bmc{F}^{\tilde{\text{F}} \vert \text{S}+\text{F}}_{t,v} + \bmc{F}^{\text{F} \vert \text{S}}_{t,v} + \bmc{F}^{\text{S}}_{t,v} .
\end{align}
According to \eqref{fisher-entropy-difference}, the additional information about the system's velocity encoded in the internal variables of the feedback controller increases dissipation, in addition to any dissipation due to the dynamics of the internal variables themselves.
This also implies that, if we ignore the hidden degrees of freedom and their dependence on the system's velocity, the dissipation of the resulting effective dynamics will underestimate the true amount of dissipation (assuming the noises acting on $\tilde{\text{F}}$ are independent from the noises acting on F and S).
Thus, even if we do not resolve the internal degrees of freedom, the entropy production rates $\bar{\sigma}_t^\text{S}$ and $\sigma_t^\text{S}$ corresponding to the dynamics \eqref{langevin-system}, respectively \eqref{langevin-sup} and \eqref{langevin-feedback}, will be lower bounds on the dissipation that is required to implement the dynamics in practice.

\subsection{Information thermodynamics}
The results of the previous section already imply that information acquired about the system by the feedback controller is related to dissipation.
We now make this relation more explicit by considering the thermodynamics of information corresponding to \eqref{fpe}.
We start by defining the mutual information between the system and the feedback controller,
\begin{align}
I_t^{\text{S}:\text{F}} = \int d\bm{r} \int d\bm{v} \int d\bm{y} \ \ln \bigg( \frac{p_t^\text{S+F}(\bm{r},\bm{v},\bm{y})}{p_t^\text{S}(\bm{r},\bm{v}) p_t^\text{F}(\bm{y})} \bigg) p_t^\text{S+F}(\bm{r},\bm{v},\bm{y}) = D_\text{KL}\big( p_t^\text{S+F} \Vert p_t^\text{S} p_t^\text{F} \big) ,
\end{align}
where $D_\text{KL}(p \Vert \tilde{p})$ denotes the Kullback-Leibler divergence or relative entropy between two probability densities $p$ and $\tilde{p}$.
As a Kullback-Leibler divergence, the mutual information is positive and vanishes only if the joint probability density factorizes $p_t^\text{S+F}(\bm{r},\bm{v},\bm{y}) = p_t^\text{S}(\bm{r},\bm{v}) p_t^\text{F}(\bm{y})$.
It therefore provides a positive measure of the correlations between the system and the feedback controller; it quantifies information contained in the joint state of S and F that is not contained in the states of S and F separately.
The mutual information is equal to the difference between the Shannon entropies of the system and the feedback controller and their joint Shannon entropy,
\begin{align}
I_t^{\text{S}:\text{F}} = S_t^\text{S} + S_t^\text{F} - S_t^\text{S+F} \label{mutual-info-shannon} .
\end{align}
Taking the time-derivative of the mutual information, we therefore have
\begin{align}
\dot{I}_t^{\text{S}:\text{F}} = \dot{S}_t^\text{S} + \dot{S}_t^\text{F} + \int d\bm{r} \int d\bm{v} \int d\bm{y} \ \ln p_t^\text{S+F}(\bm{r},\bm{v},\bm{y}) \partial_t p_t^\text{S+F}(\bm{r},\bm{v},\bm{y}) .
\end{align}
Replacing the time-derivative using \eqref{fpe} and integrating by parts, we can write this as
\begin{align}
\dot{I}_t^{\text{S}:\text{F}} = \dot{S}_t^\text{S}  &+ \int d\bm{r} \int d\bm{v} \int d\bm{y} \ \Big( \bm{j}_t^r(\bm{r},\bm{v},\bm{y}) \cdot \grad_r \ln p_t^\text{S+F}(\bm{r},\bm{v},\bm{y}) + \bm{j}_t^v(\bm{r},\bm{v},\bm{y}) \cdot \grad_v \ln p_t^\text{S+F}(\bm{r},\bm{v},\bm{y}) \Big) \\
&+ \dot{S}_t^\text{F} + \int d\bm{r} \int d\bm{v} \int d\bm{y} \ \bm{j}_t^y(\bm{r},\bm{v},\bm{y}) \cdot \grad_y \ln p_t^\text{S+F}(\bm{r},\bm{v},\bm{y}) \n .
\end{align}
The terms in the fist line depend on the time-evolution and probability currents of the system, while those in the second line depend on the feedback controller.
We therefore define the learning rates
\begin{subequations}
\begin{align}
l_t^\text{S} &= \dot{S}_t^\text{S}  + \int d\bm{r} \int d\bm{v} \int d\bm{y} \ \Big( \bm{j}_t^r(\bm{r},\bm{v},\bm{y}) \cdot \grad_r \ln p_t^\text{S+F}(\bm{r},\bm{v},\bm{y}) + \bm{j}_t^v(\bm{r},\bm{v},\bm{y}) \cdot \grad_v \ln p_t^\text{S+F}(\bm{r},\bm{v},\bm{y}) \Big), \\
l_t^\text{F} &= \dot{S}_t^\text{F} + \int d\bm{r} \int d\bm{v} \int d\bm{y} \ \bm{j}_t^y(\bm{r},\bm{v},\bm{y}) \cdot \grad_y \ln p_t^\text{S+F}(\bm{r},\bm{v},\bm{y}),
\end{align}
\end{subequations}
which quantify the contribution of the system and feedback controller, respectively, to the change in the correlations between the two, so that
\begin{align}
\dot{I}_t^{\text{S}:\text{F}} = l_t^\text{S} + l_t^\text{F} \label{learning-rates-mutual-info} .
\end{align}
Using the explicit expression of the probability currents of the system in \eqref{fpe}, we can evaluate the learning rate of the system further,
\begin{align}
l_t^\text{S} = \dot{S}_t^\text{S} + \int d\bm{r} \int d\bm{v} \int d\bm{y} \ \bigg( &\grad_r p_t^\text{S+F}(\bm{r},\bm{v},\bm{y}) \cdot \bm{v} \\
+ &\grad_v p_t^\text{S+F}(\bm{r},\bm{v},\bm{y}) \cdot \bm{m}^{-1} \Big( - \grad_r U(\bm{r}) - \bm{\gamma} \bm{v} + \bm{f}^\text{fb}(\bm{r},\bm{y}) - \bm{\gamma} \bm{m}^{-1} T \grad_v \ln p_t^\text{S+F}(\bm{r},\bm{v},\bm{y}) \Big) \bigg) \n .
\end{align}
Using the boundary conditions, the first, second and fourth term under the integral vanish, and we obtain
\begin{align}
l_t^\text{S} = \dot{S}_t^\text{S} - \text{tr} \Big( \bm{\gamma} \bm{m}^{-1} \big( T \bm{m}^{-1} \bmc{F}_{t,v}^\text{S+F} - \bm{I} \big) \Big) .
\end{align}
Using \eqref{shannon-fisher} and \eqref{fisher-entropy-difference}, we can rewrite this as
\begin{align}
l_t^\text{S} = - \sigma_t^\text{epu} - T \text{tr} \big( \bm{\gamma} \bm{m}^{-2} \bmc{F}_{t,v}^{\text{F} \vert \text{S}} \big) .
\end{align}
Comparing this to \eqref{entropy-2}, we have
\begin{align}
\sigma_t^\text{S} = \dot{S}_t^\text{S} + \sigma_t^\text{env,S} - l_t^\text{S} \geq 0 \label{second-law-info-sup} .
\end{align}
This relation is known as the second law of information thermodynamics.
While on the level of the overall dynamics of system and feedback controller, the sum of the change in the Shannon entropy and the entropy change of the environment is always positive according to \eqref{second-law}, this is not necessarily true when considering the system by itself.
However, \eqref{second-law-info-sup} states that any apparent violation of the second law in the system has to be compensated by an information exchange between the system and feedback controller, which is expressed by the learning rate $l_t^\text{S}$.
More precisely, in order for $\dot{S}_t^\text{S} + \sigma_t^\text{env,S}$ to become negative, $l_t^\text{S}$ also has to be negative, which indicates that the apparent violation of the second law is enabled by the system consuming correlations with the feedback controller.
Conversely, we also have the same relation for the feedback controller,
\begin{align}
\sigma_t^\text{F} = \dot{S}_t^\text{F} + \sigma_t^\text{env,F} - l_t^\text{F} \geq 0 .
\end{align}
Summing this with \eqref{second-law-info-sup}, we obtain
\begin{align}
\sigma_t^\text{S} + \sigma_t^\text{F} = \dot{S}_t^\text{S} + \dot{S}_t^\text{F} + \sigma_t^\text{env,S} + \sigma_t^\text{env,F} - l_t^\text{S} - l_t^\text{F} .
\end{align}
Using \eqref{learning-rates-mutual-info} and \eqref{mutual-info-shannon}, this recovers \eqref{second-law}, provided that we have
\begin{align}
\sigma_t^\text{env,S+F} = \sigma_t^\text{env,S} + \sigma_t^\text{env,F} .
\end{align}
This relation holds since we assumed the noises acting on S and F to be independent, which is equivalent to both systems being in contact with independent environments (or statistically independent parts of the same environment), so that the the entropy of the environment of S only changes due to the dynamics of S and analog for F.
Using \eqref{learning-rates-mutual-info}, we further have for the learning rate of the feedback controller,
\begin{align}
l_t^\text{F} = \dot{I}_t^\text{S:F} + \sigma_t^\text{epu} + T \text{tr} \big( \bm{\gamma} \bm{m}^{-2} \bmc{F}_{t,v}^{\text{F} \vert \text{S}} \big) .
\end{align}
This implies that the dependence of the feedback controller state on the system's velocity not only increases the overall dissipation (see \eqref{fisher-entropy-difference}) but also leads to additional information transfer from the system to the feedback controller.

\section{Feedback cooling and information thermodynamics}
We now specialize the discussion in the previous section to the case of feedback cooling.
Here, the task of the feedback controller is to reduce the overall kinetic temperature of the system, defined as
\begin{align}
T_\text{K} = \frac{\Av{\bm{v} \cdot \bm{m} \bm{v}}}{d} = \frac{\text{tr}\big( \bm{m} \bmc{V}_t \big)}{d},
\end{align}
that is, twice the kinetic energy per degree of freedom of the velocity.
Here, we recall the definition of the matrix $\bmc{V}_t$ in \eqref{velocity-covariance}.
We further focus on the steady state $p_\text{st}(\bm{r},\bm{v},\bm{y})$ of \eqref{fpe} (assuming it exists), so that all time-derivatives vanish.
For a vanishing feedback force, the steady-state of the system is the Boltzmann-Gibbs equilibrium state
\begin{align}
p_\text{eq}^\text{S}(\bm{r},\bm{v}) = \frac{1}{Z_\text{eq}^\text{S}} \exp \Bigg( - \frac{1}{T} \bigg( \frac{1}{2} \bm{v} \cdot \bm{m} \bm{v} + U(\bm{r}) \bigg) \Bigg) ,
\end{align}
where $Z_\text{eq}^\text{S}$ is the normalizing partition function.
In equilibrium, we find, as expected
\begin{align}
T_\text{K} = T,
\end{align}
in agreement with the equipartition theorem.
More generally, we have from \eqref{second-law-info-sup},
\begin{align}
\sigma_\text{st}^\text{S} = l_\text{st}^\text{F} + \frac{1}{T} \text{tr} \Big( \bm{\gamma} \bm{m}^{-1} \big( \bm{m} \bmc{V}_\text{st} - T \bm{I} \big) \Big) \geq 0 ,
\end{align}
where we used that $l_\text{st}^\text{S} = -l_\text{st}^\text{F}$ in the steady state.
If all particles have the same mass and friction coefficient, $\bm{m} = m \bm{I}$ and $\bm{\gamma} = \gamma \bm{I}$, then this reduces to the relation
\begin{align}
\sigma_\text{st}^\text{S} = l_\text{st}^\text{F} + \frac{\gamma d}{m} \bigg( \frac{T_\text{K}}{T} - 1 \bigg) \geq 0 ,
\end{align}
which implies that reducing the kinetic temperature below the environmental temperature, $T_\text{K} < T$, necessarily requires a positive learning rate of the feedback controller.
However, we note that for differing masses and friction coefficients, there is no one-to-one correspondence between the kinetic temperature and the entropy production rate.
In this case, we can in principle achieve $T_\text{K} < T$ even for vanishing entropy pumping rate.

\section{Linear model for measurement noise}
As we saw in the previous section, reducing the kinetic temperature of the particle system requires an effective feedback force that depends on the velocity of the particles, and, thereby leads to entropy pumping and information exchange between the system and feedback controller.
In practical applications, however, we usually cannot directly measure the velocity of the particles, but rather have to infer it from a measurement of their position at different times.
Moreover, the measurement of the particle's positions is generally not instantaneous and also involves errors due to measurement noise.
We model this effect by introducing a set of dynamical variables $\bm{z}(t)$, which represent to a delayed, noisy measurement of the position,
\begin{align}
\bm{z}(t) = \frac{1}{\tau} \int_0^t ds \ e^{-\frac{t-s}{\tau}} \big( \bm{r}(s) + \sqrt{2 D} \bm{\eta}(s) \big) \label{measurement-variable} ,
\end{align}
where $\bm{\eta}(t)$ is a vector of independent Gaussian white noises representing the measurement noise on the individual coordinates.
It is clear that, in the absence of measurement noise, $D = 0$, $\bm{z}(t)$ most accurately represents $\bm{r}(t)$ for $\tau \rightarrow 0$, that is, when the delay is as short as possible.
However, the delay also has the effect of reducing the effective measurement noise by time-averaging and therefore a finite delay (and thus averaging time) will lead to a more accurate result for the position, as long as the averaging time is short compared to the characteristic time scales of the system.
Taking $\bm{z}(t)$ as the measured result of the position, we then estimate the velocity via its derivative, $\bm{u}(t) = \dot{\bm{z}}(t)$, and use this estimate to apply feedback to the system.
In the simplest case, this feedback force is linear in the measured velocity
\begin{align}
\bm{f}^\text{fb}(\bm{u}) = - \gamma^\text{fb} \bm{u} \label{linear-feedback} .
\end{align}
We note however, that this description differs from the general model in \eqref{fpe}, since the feedback force now depends on the derivative of the feedback controller's degrees of freedom.
Specifically the equations of motion of the system and feedback controller are (assuming identical masses and friction coefficients)
\begin{subequations}
\begin{align}
\dot{\bm{r}}(t) = \bm{v}(t), \qquad m \dot{\bm{v}}(t) &= - \grad_r U(\bm{r}(t)) - \gamma \bm{v}(t) - \frac{\gamma^\text{fb}}{\tau} \big( \bm{r}(t) - \bm{z}(t) \big) - \sqrt{\frac{2 (\gamma^\text{fb})^2 D}{ \tau^2}} \bm{\eta}(t) + \sqrt{2 \gamma T} \bm{\xi}(t) \\
\dot{\bm{z}}(t) &= \frac{1}{\tau} \Big( - \big( \bm{z}(t) - \bm{r}(t) \big) + \sqrt{2 D} \bm{\eta}(t) \Big)  . 
\end{align} \label{langevin-measurement}%
\end{subequations}
where the second line follows by differentiating \eqref{measurement-variable} and we replaced $\dot{\bm{z}}(t)$ in the first line with the latter's equation of motion.
This describes the dynamics of the degrees of freedom of the system $(\bm{r},\bm{v})$ and the feedback controller $\bm{z}$, similar to \eqref{langevin} and \eqref{langevin-feedback}, however, the noises affecting the system and feedback controller are now no longer independent, since the feedback force feeds the measurement noise directly back into the system.
From a mathematical point of view, the diffusion matrix corresponding to \eqref{langevin-measurement} is not diagonal.
As discussed in Ref.~\cite{Che19} this corresponds to a non-bipartite structure and we can no longer clearly separate the contributions of the system and feedback controller to the dissipation.
One way to avoid this issue is to introduce a short but finite correlation time $\tau^\text{noise}$ for the measurement noise,
\begin{align}
\dot{\bm{\eta}}(t) = - \frac{1}{\tau^\text{noise}} \big( \bm{\eta}(t) - \tilde{\bm{\xi}}(t) \big) ,
\end{align}
where $\tilde{\bm{\xi}}(t)$ is Gaussian white noise.
The correlation function of the measurement noise is then
\begin{align}
\Av{\eta_k(t) \eta_j(t')} = \frac{1}{2 \tau^\text{noise}} \delta_{kj} e^{-\frac{|t-t'|}{\tau^\text{noise}}},
\end{align}
which reduces to white noise in the limit $\tau^\text{noise} \rightarrow 0$.
We then have the equations of motion,
\begin{subequations}
\begin{align}
\dot{\bm{r}}(t) = \bm{v}(t), \qquad m \dot{\bm{v}}(t) &= - \grad_r U(\bm{r}(t)) - \gamma \bm{v}(t) - \gamma^\text{fb} \bm{u} + \sqrt{2 \gamma T} \bm{\xi}(t) \\
\tau \dot{\bm{z}}(t) &= - \big( \bm{z}(t) - \bm{r}(t) \big) + \sqrt{2 D} \bm{\eta}(t) \\
\tau \dot{\bm{u}}(t) &= - \big( \bm{u}(t) - \bm{v}(t) \big) + \frac{\sqrt{2 D}}{\tau^\text{noise}} \big( - \bm{\eta}(t) + \tilde{\bm{\xi}}(t) \big) \\
\tau^\text{noise} \dot{\bm{\eta}}(t) &= - \bm{\eta}(t) + \tilde{\bm{\xi}}(t)  . 
\end{align} \label{langevin-measurement-corr}%
\end{subequations}
Treating $\bm{y} = (\bm{z},\bm{u},\bm{\eta})$ as the degrees of freedom of the feedback controller, we see that the diffusion matrix of the feedback controller itself becomes non-diagonal, however, we can clearly separate between the noises $\bm{\xi}$ acting on the system and $\tilde{\bm{\xi}}$ acting on the feedback controller.
Moreover, since only $\bm{u}$ directly enters the feedback force, we can define the effective feedback force and excess information flow by considering only the conditional probability density of $\bm{u}$, $p_t^{\text{F} \vert \text{S}}(\bm{u} \vert \bm{r}, \bm{v})$.
In the language of Eq.~(18) of the main text, $\bm{u}$ corresponds to the direct degrees of freedom, while $\bm{z}$ and $\bm{\eta}$ are indirect degrees of freedom that are required for the description of the dynamics of the feedback controller but do not directly impact the feedback force.

We now focus on the case of linear interactions between the particles $\bm{U}(\bm{r}) = \frac{1}{2} (\bm{r} - \bm{r}_0) \cdot \bm{K} (\bm{r} - \bm{r}_0)$, where $\bm{K}$ is a positive definite matrix of coupling constants.
Since, for this choice, \eqref{langevin-measurement-corr} is a linear set of equations in the involved variables, we can compute its solution explicitly.
Specifically, introducing the vector $\bm{w} = (\bm{r},\bm{v},\bm{z},\bm{u},\bm{\eta})$, we can write \eqref{langevin-measurement-corr} as
\begin{gather}
\dot{\bm{w}}(t) = - \bmc{K} \big(\bm{w}(t) - \bm{w}_0 \big) + \bmc{G} \bm{\theta}(t) \qquad \text{with} \\
\bmc{K} = \begin{pmatrix} 
0 & -\bm{I} & 0 & 0 & 0 \\ \bm{K}/m & \gamma/m \bm{I} & 0 & \gamma^\text{fb}/m \bm{I} & 0 \\
- 1/\tau \bm{I} & 0 & 1/\tau \bm{I} & 0 & \sqrt{2 D} \bm{I} \\
0 & - 1/\tau \bm{I} & 0 & 1/\tau \bm{I} & -\sqrt{2 D}/\tau \bm{I} \\
0 & 0 & 0 & 0 & 1/\tau^\text{noise} \bm{I} 
\end{pmatrix} \qquad
\bmc{G} = \begin{pmatrix}
0 & 0 & 0 & 0 & 0 \\
0 & \sqrt{2 \gamma T}/m \bm{I} & 0 & 0 & 0\\
0 & 0 & 0 & 0 & 0 \\
0 & 0 & 0 & 0 & \sqrt{2 D}/\tau^\text{noise} \bm{I} \\
0 & 0 & 0 & 0 & 1/\tau^\text{noise} \bm{I} 
\end{pmatrix} \qquad 
\bm{w}_0 = \begin{pmatrix}
\bm{r}_0 \\ 0 \\ \bm{r}_0 \\ 0 \\ 0
\end{pmatrix} \n ,
\end{gather}
where $\bm{\theta}(t)$ is a vector of independent Gaussian white noises.
The steady state of the joint system is then the Gaussian probability density
\begin{align}
p^{\text{F} + \text{S}}_\text{st}(\bm{w}) = \frac{1}{(2 \pi)^{5d/2} \det(\bm{\Xi})} \exp \bigg( - \frac{1}{2} (\bm{w} - \bm{w}_0) \cdot \bm{\Xi}^{-1} (\bm{w} - \bm{w}_0) \bigg),
\end{align}
where the covariance matrix $\bm{\Xi}$ is obtained by solving the Lyapunov equation
\begin{align}
\bmc{K} \bm{\Xi} + \bm{\Xi} \bmc{K}^\text{T} = \bmc{G} \bmc{G}^\text{T} \label{lyapunov} .
\end{align}
Marginalizing over $\bm{z}$ and $\bm{\eta}$ yields the joint probability density of the system state and $\bm{u}$, $p^{\text{F} + \text{S}}_\text{st}(\bm{r},\bm{v},\bm{u})$ whose covariance matrix we denote by $\bm{\Phi}$, which is obtained from $\bm{\Xi}$ by removing the rows and columns corresponding to $\bm{z}$ and $\bm{\eta}$.
The conditional probability density $p^{\text{F} \vert \text{S}}_\text{st}(\bm{u} \vert \bm{r},\bm{v})$ of $\bm{u}$ is likewise Gaussian with mean $\hat{\bm{u}}(\bm{r},\bm{v})$ and covariance matrix $\hat{\bm{\Phi}}^{u}$
\begin{align}
\hat{\bm{u}}(\bm{r},\bm{v}) = \bm{\Phi}^{u,rv} \big(\bm{\Phi}^{rv,rv} \big)^{-1} \begin{pmatrix} \bm{r} - \bm{r}_0 \\ \bm{v} \end{pmatrix}, \qquad \text{and} \qquad \hat{\bm{\Phi}}^u = \bm{\Phi}^{u,u} - \bm{\Phi}^{u,rv} \big(\bm{\Phi}^{rv,rv}  \big)^{-1} \bm{\Phi}^{rv,u},
\end{align}
where we defined the blocks of the covariance matrix $\bm{\Phi}$ as
\begin{align}
\bm{\Phi} = \begin{pmatrix}
\bm{\Phi}^{rv,rv} & \bm{\Phi}^{rv,u} \\ \bm{\Phi}^{u,rv} &\bm{\Phi}^{u,u} 
\end{pmatrix} .
\end{align}
The effective feedback force is then obtained by replacing $\bm{u}$ with the conditional average $\hat{\bm{u}}(\bm{r},\bm{v})$ in \eqref{linear-feedback},
\begin{align}
\bar{\bm{f}}^\text{fb}(\bm{r},\bm{v}) = - \gamma^\text{fb} \hat{\bm{u}}(\bm{r},\bm{v}) = - \gamma^\text{fb} \bm{\Phi}^{u,rv} \big(\bm{\Phi}^{rv,rv} \big)^{-1} \begin{pmatrix} \bm{r} - \bm{r}_0 \\ \bm{v} \end{pmatrix} \label{linear-feedback-effective} .
\end{align}
Even though we consider $\bm{u}$ as an estimate of the velocity $\bm{v}$ of the system, the effective feedback force generally depends on the position of the system as well.
This dependence arises from the finite-time measurement process \eqref{measurement-variable}, which causes the derivative of the measurement $\bm{u}$ to deviate from the instantaneous velocity $\bm{v}$.
Using \eqref{linear-feedback-effective}, we obtain the entropy pumping rate as
\begin{align}
\sigma^\text{epu}_\text{st} = - \frac{1}{m} \Av{\grad_v \cdot \bar{\bm{f}}^\text{fb}} = - \frac{\gamma^\text{fb}}{m} \text{tr}\bigg( \bm{\Phi}^{u,rv} \Big[\big(\bm{\Phi}^{rv,rv} \big)^{-1} \Big]^{rv,v} \bigg) .
\end{align}
The deviation of the feedback force \eqref{linear-feedback} from the effective value \eqref{linear-feedback-effective} is given by the covariance matrix of the conditional probability density,
\begin{align}
\Av{\Vert \delta \bm{f}^\text{fb} \Vert^2} = \Av{\Vert \bm{f}^\text{fb} - \bar{\bm{f}}^\text{fb} \Vert^2} = (\gamma^\text{fb})^2 \text{tr} \big(\hat{\bm{\Phi}}^u\big) ,
\end{align}
while the excess information flow is computed as
\begin{align}
l_\text{st}^\text{ex} = \frac{\gamma T}{m^2} \Av{ \Vert \grad_v \ln p^{\text{F} \vert \text{S}}_\text{st} \Vert^2} = \frac{\gamma T}{m^2} \text{tr} \bigg( \Big[ \big(\bm{\Phi}^{rv,rv} \big)^{-1} \bm{\Phi}^{rv,u} \big(\hat{\bm{\Phi}}^u \big)^{-1} \bm{\Phi}^{u,rv} \big(\bm{\Phi}^{rv,rv} \big)^{-1} \Big]^{v,v} \bigg) .
\end{align}
The kinetic temperature is given by the $\bm{v}$-component of the covariance matrix $\bm{\Phi}$,
\begin{align}
T^\text{K} = \frac{m}{d} \text{tr}\big(\bm{\Phi}^{v,v} \big) .
\end{align}

For concrete expressions, we consider the one-dimensional case with $U(r) = \frac{1}{2} k r^2$,
\begin{subequations}
\begin{align}
\dot{r}(t) = v(t), \qquad m \dot{v}(t) &= - k r(t) - \gamma v(t) - \gamma^\text{fb} u(t) + \sqrt{2 \gamma T} \xi(t) \\
\tau \dot{z}(t) &= - \big( z(t) - r(t) \big) + \sqrt{2 D} \eta(t) \\
\tau \dot{u}(t) &= - \big( u(t) - v(t) \big) + \frac{\sqrt{2 D}}{\tau^\text{noise}} \big( - \eta(t) + \tilde{\xi}(t) \big) \\
\tau^\text{noise} \dot{\eta}(t) &= - \eta(t) + \tilde{\xi}(t)  . 
\end{align} %
\end{subequations}
We can bring these equations into dimensionless form by introducing the dimensionless variables and parameters
\begin{gather}
\nu(t) = \frac{v(t)}{v^\text{th}}, \qquad \rho(t) = \frac{r(t)}{v^\text{th} \tau^\text{th}}, \qquad \alpha = \frac{\gamma^\text{fb}}{\gamma}, \qquad \Omega = \frac{\sqrt{k m}}{\gamma} \\
\mu(t) = \frac{u(t)}{v^\text{th}}, \quad \chi(t) = \frac{z(t)}{v^\text{th} \tau^\text{th}}, \quad \epsilon(t) = \sqrt{\tau^\text{th}} \eta(t), \quad \theta = \frac{\tau}{\tau^\text{th}}, \quad \bar{\theta} = \frac{\tau^\text{noise}}{\tau^\text{th}}, \quad \mathcal{D} = \frac{D}{(v^\text{th})^2 (\tau^\text{th})^3} = \frac{\gamma^3 D}{m^2 T} \n,
\end{gather}
where we defined $\tau^\text{th} = m/\gamma$ and $v^\text{th} = \sqrt{T/m}$.
We obtain
\begin{subequations}
\begin{align}
\dot{\rho}(s) = \nu(s), \qquad \dot{\nu}(s) &= - \Omega^2 \rho(s) - \nu(s) - \alpha \mu(s) + \sqrt{2} \xi(s) \\
\theta \dot{\chi}(s) &= - \big( \chi(s) - \rho(s) \big) + \sqrt{2 \mathcal{D}} \epsilon(t) \\
\theta \dot{\mu}(s) &= - \big( \mu(s) - \nu(s) \big) + \frac{\sqrt{2 \mathcal{D}}}{\bar{\theta}} \big( - \epsilon(s) + \tilde{\xi}(s) \big) \\
\bar{\theta} \dot{\epsilon}(s) &= - \epsilon(s) + \tilde{\xi}(s)   
\end{align} \label{langevin-measurement-corr-1D}%
\end{subequations}
with the dimensionless time $s = t/\tau^\text{th}$.
Here, $\xi(s)$ and $\tilde{\xi}(s)$ are independent dimensionless Gaussian white noises with correlation function $\Av{\xi(s) \xi(s')} = \delta(s-s')$.
In terms of the dimensionless parameters defined above, the ratio of the kinetic and environmental temperature is given by
\begin{align}
\frac{T^\text{K}}{T} = \frac{1 + \theta \big( 1 + \alpha + \theta \Omega^2 \big) +  \alpha^2 \mathcal{D} \frac{  1 + \alpha + (\theta + \bar{\theta}) \Omega^2 }{\theta + \bar{\theta} + \theta \bar{\theta} + \bar{\theta}^2 ( 1 + \alpha + (\theta + \bar{\theta} ) \Omega^2 )}}{1 + \theta + \alpha (1 + \theta) + \theta^2 \Omega^2} \label{kinetic-temperature} .
\end{align}
This result is obtained by solving the Lyapunov equation \eqref{lyapunov} and then evaluating the $v$-component of the covariance matrix, which defines the kinetic temperature $T^\text{K} = m \Av{v^2}$.
While the precise dependence of this expression on the parameters is complicated, we can nevertheless understand its general behavior.
In the absence of measurement noise, $\mathcal{D} = 0$, the kinetic temperature is minimized in the limit of strong feedback $\alpha \rightarrow \infty$, and its minimal value is $T^\text{K}/T = \theta/(1+\theta)$.
Thus, no cooling is possible if the measurement time $\tau$ is long compared to the thermalization time $\tau^\text{th}$ ($\theta \gg 1$).
On the other hand, in the limit of an instantaneous noise-free measurement $\theta \rightarrow 0$, we can in principle cool to zero temperature.
However, for any finite measurement noise $\mathcal{D} > 0$, the second term in the numerator will dominate for sufficiently large $\alpha$, causing the temperature to increase with increasing feedback strength.
The reason is that, since the feedback force is influenced by the measurement noise, applying a strong feedback force essentially corresponds to feeding back the measurement noise into the system, thereby heating it.
We also note that the kinetic temperature simplifies in the limit $\bar{\theta} \rightarrow 0$, which corresponds to white measurement noise
\begin{align}
\frac{T^\text{K}}{T} = \frac{1 + \big( \theta + \frac{\alpha^2 \mathcal{D}}{\theta} \big) \big( 1 + \alpha + \theta \Omega^2 \big) }{1 + \theta + \alpha (1 + \theta) + \theta^2 \Omega^2} \label{kinetic-temperature-white} .
\end{align}
This expression shows that, for finite measurement noise, a finite integration time $\theta$ is necessary for cooling, since, otherwise, the fluctuations due to the measurement noise will dominate the measured signal.
For given values of the remaining parameters, the kinetic temperature, \eqref{kinetic-temperature} or \eqref{kinetic-temperature-white}, has a unique minimum value as a function of $\alpha$ (corresponding to the magnitude of the feedback force) and $\theta$ (corresponding to the integration time in the measurement).
Due to the non-linear dependence on the parameters, we determine this minimal value using Mathematica's \texttt{NMinimize} command.
We remark that, as the system is one-dimensional, we can write the effective feedback force as
\begin{align}
\bar{f}^\text{fb}(r,v) = -\gamma^\text{fb} \big(c^r r + c^v v \big)
\end{align}
with appropriately defined constants $c^r$ and $c^v$.
Likewise, the conditional covariance matrix of $u$ reduces to a scalar $\hat{\Phi}^u$.
Thus, the entropy pumping rate, excess information flow and deviations of the feedback force can be written as
\begin{align}
\sigma_\text{st}^\text{epu} = \frac{\gamma^\text{fb} c^v}{m}, \quad l_\text{st}^\text{ex} = \frac{\gamma T}{m^2 \hat{\Phi}^u} , \quad \Av{\vert f^\text{fb} - \bar{f}^\text{fb} \vert^2} = (\gamma^\text{fb} c^v)^2 \hat{\Phi}^u .
\end{align}
These satisfy the identity
\begin{align}
\gamma T \big(\sigma_\text{st}^\text{epu} \big)^2 = l_\text{st}^\text{ex} \Av{\vert f^\text{fb} - \bar{f}^\text{fb} \vert^2},
\end{align}
that is, the trade-off Eq.~(7) of the main text is satisfied with equality.
Moreover, as the dependence of the feedback force on $r$ can be interpreted as a renormalization of the trapping potential, the feedback force is essentially proportional to the velocity.
As shown in the main text, this also leads to equality in Eq.~(4) of the main text.
In summary, the cooling scheme above simultaneously maximizes the cooling efficiency and the entropy pumping efficiency,
\begin{align}
\eta^\text{cool} = \frac{\gamma \big(\frac{T}{T^\text{K}} - 1 \big)}{m \sigma^\text{epu}_\text{st}} = 1, \qquad \eta^\text{epu} = \frac{\sigma_\text{st}^\text{epu}}{\sigma^\text{epu}_\text{st} + l^\text{ex}_\text{st}} = \frac{1}{1 + \frac{\gamma T \sigma_\text{st}^\text{epu}}{\Av{\vert \delta f^\text{fb} \vert^2}}},
\end{align}
where the rightmost expressions are upper bounds in the generic case.

\end{widetext}

\end{document}